\documentclass[reprint, prb, aps, floatfix, nofootinbib, superscriptaddress]{revtex4-2}
\usepackage{amsmath}
\usepackage{amsfonts}
\usepackage{rotating}
\usepackage{graphicx}
\usepackage{chemformula}
\usepackage{mathtools}
\usepackage{tabularx}
\usepackage{color}
\usepackage{bm}
\usepackage{dcolumn}

\usepackage{soul}

\usepackage[normalem]{ulem}
\usepackage{cancel}
\definecolor{purple}{RGB}{160, 32, 240}

\renewcommand{\mathbf}[1]{\bm{#1}}

\usepackage{hyperref}
\hypersetup{
    colorlinks=true,
    linkcolor=blue,
    citecolor=blue,
    urlcolor=blue
}
\begin{document}

\title{Magnetic Phase Diagrams and Spin Hamiltonian of \\Monoclinic $\alpha$-RuCl$_3$ from Angle-Dependent Torque Studies}

\author{Daniel Antoniou}
\affiliation{Clarendon Laboratory, University of Oxford Physics Department, Parks Road, Oxford OX1 3PU, UK}
\author{Danrui Ni}
\affiliation{Department of Chemistry, Princeton University, Princeton, NJ 08544, USA}
\author{John S.~Pearce}
\affiliation{Clarendon Laboratory, University of Oxford Physics Department, Parks Road, Oxford OX1 3PU, UK}
\author{Robert J.~Cava}
\affiliation{Department of Chemistry, Princeton University, Princeton, NJ 08544, USA}
\author{Amalia I.~Coldea}
\affiliation{Clarendon Laboratory, University of Oxford Physics Department, Parks Road, Oxford OX1 3PU, UK}
\author{Radu Coldea}
\affiliation{Clarendon Laboratory, University of Oxford Physics Department, Parks Road, Oxford OX1 3PU, UK}

\begin{abstract}
The layered honeycomb $\alpha$-RuCl$_3$ has been much explored as a candidate to display unconventional cooperative magnetism between spin-orbit entangled $j_{\rm eff}=1/2$ Ru$^{3+}$ magnetic moments stabilized by strongly-frustrated, Kitaev and other bond-dependent anisotropic interactions. Most studies so far have focused on relatively large single crystals that undergo a structural phase transition from monoclinic to rhombohedral upon cooling below 150~K and a magnetic transition to zigzag order around 7~K. Here we study the magnetism of high-quality, very small (sub 100~$\mu$m diameter) single crystals, which remain monoclinic upon cooling to low temperatures and display a sharp magnetic transition at 14~K. Using highly-sensitive piezo-cantilever magnetic torque measurements in fields up to 16~T, we report a comprehensive study of the magnetic phase diagrams for magnetic field rotated in three orthogonal crystallographic planes. Our extensive torque data collected upon varying almost continuously the angular orientation of the field as well as the field magnitude, allows clear detection of phase transitions via anomalies in the raw torque data and its higher order derivatives with respect to both angle and field magnitude. The obtained magnetic phase diagrams show many differences compared to rhombohedral samples, displaying reduced rotational symmetry, substantially higher fields required to suppress the spontaneous magnetic order, and distinct field-induced phases. We provide a direct comparison of the torque data with mean-field calculations for magnetic Hamiltonians appropriate for $\alpha$-RuCl$_3$ allowing for the monoclinic lattice symmetry and propose minimal magnetic structure models for all explored regions of the magnetic phase diagrams.
\end{abstract}

\date{\today}
\maketitle
\section{Introduction}
\begin{figure}[tb]
\centering
\includegraphics[scale=1]{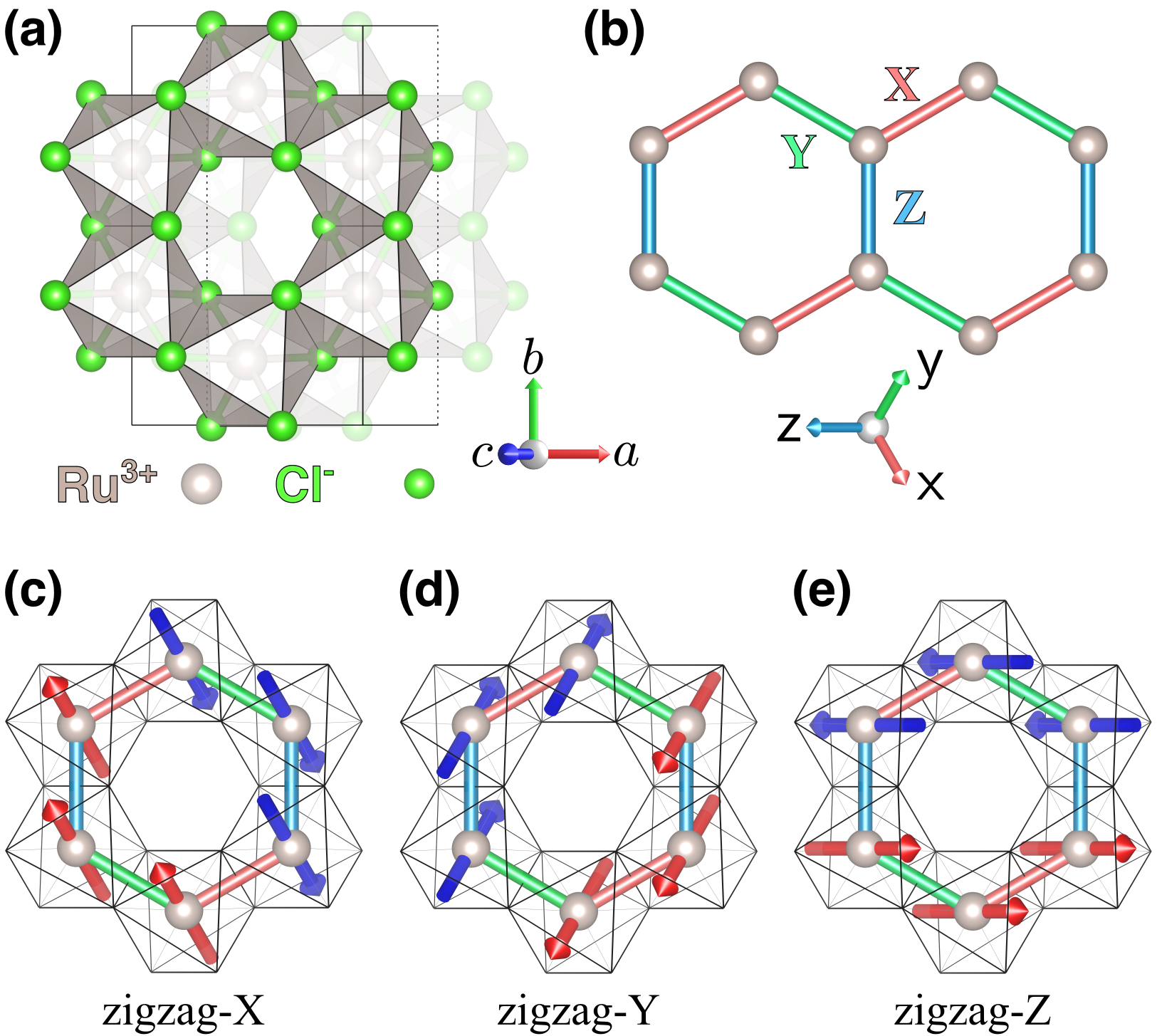}
\caption{(a) Monoclinic crystal structure of $\alpha$-RuCl$_3$ projected onto the $ab$ plane, where the thin lines show the unit cell outline. (b) Schematic of the Ru honeycomb in the $ab$ plane showing the X/Y/Z bonds (color-coded red/green/blue) with Kitaev exchange along the (cubic) ${\sf x}/{\sf y}/{\sf z}$ axes, respectively, shown in the lower triad. (c,d,e) Schematic of the zigzag-X, -Y and -Z antiferromagnetic order patterns discussed in the text.}\label{F:C2m}
\end{figure}

Quantum magnets with strong spin-orbit coupling can have anisotropic exchange interactions that can become strongly frustrated on some lattice geometries and stabilize unconventional cooperative magnetic behavior. This could range from non-trivial forms of magnetic order to long-sought quantum spin liquids, such as predicted for the Kitaev model on the honeycomb lattice with Ising exchanges on every bond, but with the Ising axes mutually orthogonal for the three bonds that meet at every lattice site, as illustrated by the bond color coding in Fig.~\ref{F:C2m}(b). In the van der Waals magnet $\alpha$-RuCl$_3$, octahedrally-coordinated Ru$^{3+}$ (4$d^5$) ions have been proposed to have $j_{\rm eff}=1/2$ magnetic moments with mixed spin-orbital character, with a dominant superexchange interaction between nearest-neighbors of Kitaev form \cite{Sears2015}. Key experimental results include the observation of a complex spectrum of excitations containing strong scattering continua indicative of important quantum fluctuation effects \cite{Banerjee2017,Ran2017}, and also unconventional thermal Hall physics in applied magnetic field \cite{Yokoi2021,Czajka2021,Bruin2022}. The interpretation of these results is still actively debated, for a review see Ref.~\cite{Kee2023}. There is consensus that (ferromagnetic) Kitaev interactions are indeed dominant and are integral to explaining some of the experimental results. However, there is clear evidence that the spin Hamiltonian also contains other substantial exchange terms, which cannot be neglected, such as Heisenberg exchange and other bond-dependent anisotropic interactions. These terms are essential to explain the stability of the observed long-range zigzag magnetic order in zero field and also to reproduce features of the phase diagram in applied magnetic field as discussed extensively in Ref.~\cite{Moeller2025}.

Characterizing the relative magnitude of the different sources of anisotropy is essential for developing a microscopic understanding of the observed complex magnetic behavior. The motivation behind our present experimental study is to probe in detail the magnetic anisotropy by performing a comprehensive characterization of the magnetic behavior for applied magnetic field in three orthogonal crystallographic planes. We use highly-sensitive angle-dependent torque magnetometry on high-quality un-twinned single crystals, and compare directly the observed angle- and field magnitude-dependence of the observed torque with calculations for anisotropic exchange Hamiltonians.

It is well established that at room temperature the honeycomb layers in $\alpha$-RuCl$_3$ are stacked vertically with an in-plane offset leading to an overall monoclinic $C2/m$ crystal structure illustrated in Fig.~\ref{F:C2m}(a). However, the low temperature crystal and magnetic structure is different depending on the crystal size. Large (mm-size) bulk single crystals undergo a first order structural phase transition upon cooling below 150~K \cite{Park2024,Kim2024}, when the honeycomb layers shift in-plane and re-arrange into a rhombohedral $R\bar{3}$ structure, with magnetic ordering in the form of a zigzag antiferromagnetic pattern developing below 7~K. Most experimental studies so far, including torque magnetometry \cite{Froude-Powers2024}, have focused on such large samples. On the other hand, powder samples made of small polycrystalline grains do not undergo a structural transition upon cooling, but remain monoclinic down to the lowest temperatures probed and display a transition to zigzag antiferromagnetic order near 13–14~K \cite{Johnson2015,Banerjee2016}. To provide insight into how subtle changes in the crystal structure lead to marked differences in the magnetic behavior we have performed comprehensive torque magnetometry studies on high-quality un-twinned small crystals (less than 100$\mu$m diameter) that remain monoclinic down to the lowest temperatures and undergo a sharp magnetic transition at $T_{\rm N}=14$~K. We find that the magnetic phase diagrams differ markedly from those of rhombohedral samples, displaying reduced rotational symmetry and requiring substantially higher magnetic fields to suppress the spontaneous magnetic long range order. We relate those effects to changes in the bond-dependent anisotropic exchange couplings between the rhombohedral and monoclinic samples.

The high structural ordering of our samples is confirmed by the observation of sharp diffraction peaks with no detectable diffuse scattering, indicating minimal, if any, layer stacking faults.
Magnetic phase transitions appear as very sharp features in our torque measurements as a function of varying temperature, field magnitude or orientation, again consistent with minimal, if any, structural disorder effects. Hence, we believe our results are representative of samples that have a high degree of structural order, but which do not undergo the monoclinic to rhombohedral structural transition possibly because of strain effects associated with the small sample size. We note that artificially grown samples of ultrathin (few-layers) $\alpha$-RuCl$_3$ intercalated between graphite layers also remain monoclinic down to base temperatures \cite{Massicotte2024} with a magnetic transition at the same $T_{\rm N}=14$~K and the deduced temperature-magnetic field phase diagrams for field along $a$ and $b$ show many similarities to our current work. Our studies are also complementary to resonant torsion magnetometry studies (which probe not the torque, but its derivative with respect to the applied field angle) on small crystals of $\alpha$-RuCl$_3$ \cite{Modic2018,Nasir2026}. All these studies suggest that samples in the thin and ultrathin limit show a common behavior, distinct from the large, bulk single crystals.

The remainder of this paper is organized as follows. Section~\ref{S:expdetails} gives details of the experimental setup used to perform angle-dependent piezo-cantilever torque measurements. The following Sec.~\ref{S:results} presents \mbox{temperature-,} angle- and field magnitude-dependence of the magnetic torque on un-twinned single-crystals for external field applied in three orthogonal planes, $bc^*$, $ac^*$ and $ab$, where labels refer to the monoclinic crystallographic axes. We first analyze the temperature-dependence of the magnetic anisotropy in the three crystallographic planes, then present a systematic analysis of the anomalies seen in the raw torque signal or its higher order derivatives as a function of temperature, angle or field magnitude and then summarize the obtained magnetic phase diagrams. Sec.~\ref{S:calculations} presents mean-field calculations for minimal spin Hamiltonians for a two-dimensional honeycomb lattice including Kitaev and other anisotropic exchanges inspired by current models proposed for $\alpha$-RuCl$_3$ and adapted to allow for non-equivalent bonds in the honeycomb lattice due to the monoclinic crystal lattice symmetry, as proposed in \cite{Cen2025}. For such models three antiferromagnetic zigzag orders, so called zigzag-X, -Y and -Z, and paramagnetic field-polarized (P) are expected in different parts of the magnetic phase diagrams. Sec.~\ref{S:comparison} presents a direct comparison of the observed torque data with the model calculations. The predicted characteristic signatures in torque upon crossing boundaries between the above four phases are clearly observed in our experiments, providing a sound basis to assign the form of the two-dimensional magnetic order in all probed regions of the magnetic phase diagrams. In Sec.~\ref{S:R-3} we compare and contrast the observed magnetic phase diagrams with those reported for rhombohedral samples. Finally, Sec.~\ref{S:conclusions} summarizes the main results and conclusions.

The three Appendices present structural characterization of the measured samples, technical details of the theoretical calculations and for completeness, torque measurements on a two-twin sample. Appendix~\ref{A:xrays} reports X-ray diffraction measurements that confirm the monoclinic $C2/m$ crystal structure with minimal stacking faults and no structural transition upon cooling. Appendix~\ref{A:calculations} gives technical details of the numerical mean-field calculations of the magnetic phase diagrams for model spin Hamiltonians. Appendix~\ref{A:twins} presents torque measurements on a sample containing two almost equal-weight monoclinic twins related by a 120$^\circ$ rotation around a common $c^*$ axis and explains how to rationalize the observed very complex phase diagram and torque angular dependence based on a superposition of the responses of the two monoclinic twins.

\section{Experimental details}\label{S:expdetails}
Crystals of $\alpha$-RuCl$_3$ were synthesized using a chemical vapor transport (CVT) method which produced thin plate crystals with natural faceted edges. Commercially purchased polycrystalline $\alpha$-RuCl$_3$ (Alfa Aesar, anhydrous) was purified and crystallized via CVT with no external transport agent. The temperature at the hot and cold ends were set to 750 and 650$^\circ$C, respectively. Several sub-100$\mu$m diameter samples suitable for torque measurements were mechanically extracted from the growth batch and screened using X-ray diffraction to find un-twinned single crystals and details are given in Appendix~\ref{A:xrays}. All samples used for the torque measurements have the monoclinic $C2/m$ structure illustrated in Fig.~\ref{F:C2m}(a). In the following we use the set of orthogonal axes $a$, $b$ (along the shortest Ru-Ru bond in the honeycomb layers) and $c^*$ (normal to the honeycomb layers) to specify the magnetic field orientation with respect to the crystallographic axes.

\begin{figure}[tb]
\centering
\includegraphics[scale=1]{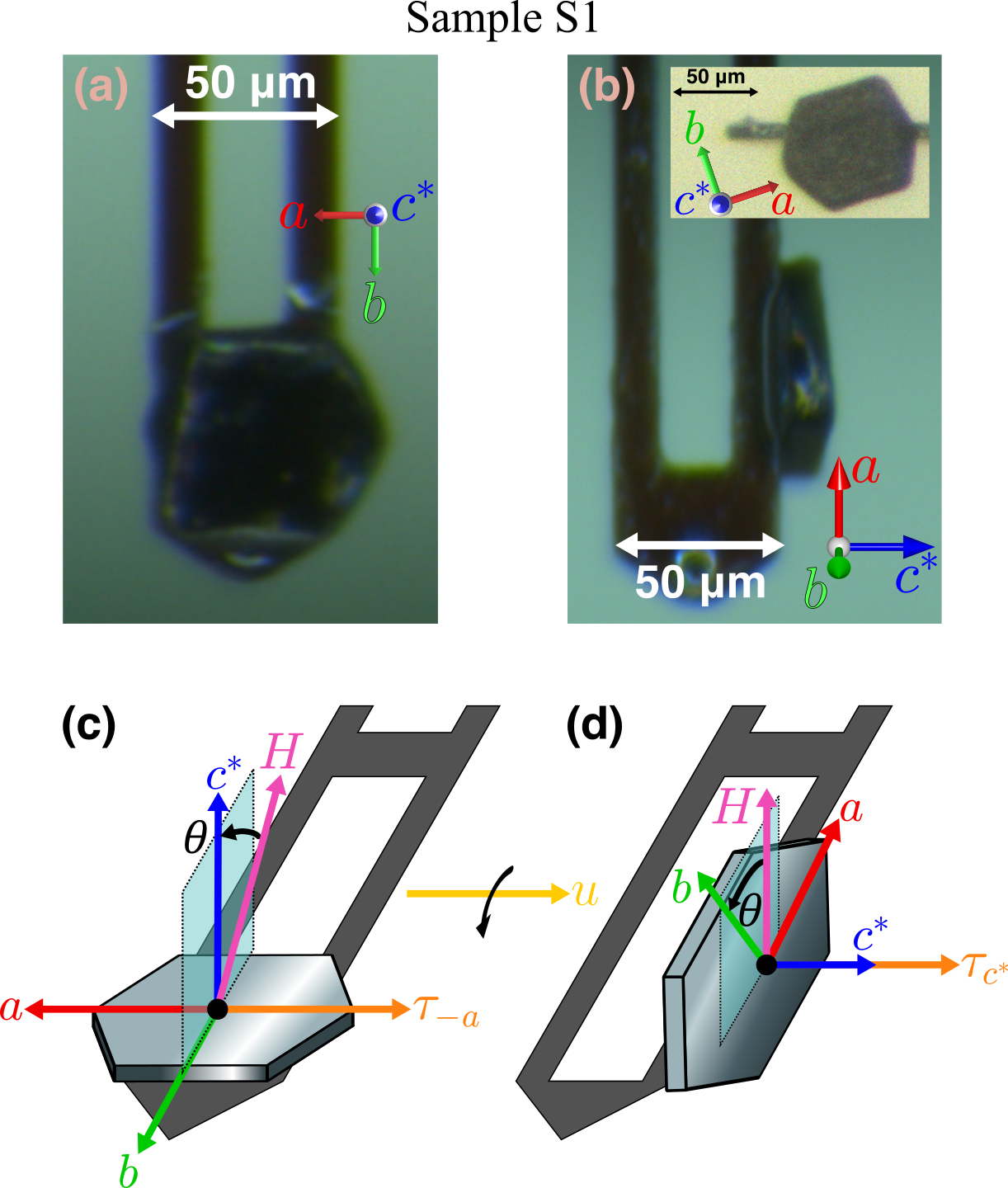}
\caption{(Top) Representative photographs of sample S1, mounted for torque measurements for field in the $bc^*$ (left) and $ab$ (right) planes, with the reference crystallographic $\bm{a}$, $\bm{b}$ and $\bm{c^*}$ axes indicated by colored arrows. (Bottom) Corresponding schematic diagram of the piezocantilevers with the sample represented by a thin elongated hexagonal platelet, mounted on the top/side of the cantilever platform in the left/right panels. The transparent light-blue shaded vertical rectangle in the bottom panels indicates the plane of rotation, with the rotation angle $\theta$ defined (up to an offset) as the angle between a reference crystallographic axis in this plane and the (fixed) applied field $\bm{H}$. The horizontal orange arrows labeled $\tau_u$, with $\bm{u}$ (yellow arrow) along the cantilever bending axis, indicate the torque component measured in the experiments.}
\label{F:levers}
\end{figure}

For the torque measurements the samples were mounted with vacuum grease on Seiko PRC400 piezo-cantilevers installed on a rotating platform inside a 16~T Quantum Design PPMS with a base temperature of 2~K. The platform rotation axis was parallel to the cantilever bending axis $\bm{u}$ and normal to the (fixed) direction of a vertically-applied magnetic field $\bm{H}$, as illustrated in Fig.~\ref{F:levers}(c-d). In the presence of the applied field a sample of magnetic moment $\bm{m}$ is subject to a torque $\bm{\tau}=\bm{m}\times \mu_0\bm{H}$. The torque component along the cantilever bending axis $\bm{u}$, $\tau_u=\bm{\tau} \cdot \bm{u}$ bends the cantilever creating mechanical stress at its base, detected in a Wheatstone bridge circuit via a voltage change $\Delta V$, which varies linearly with $\tau_u$ in the small $\tau_u$ regime.

To obtain the angular dependence of the torque $\tau_u$, the cantilever platform was rotated over an almost 360$^{\circ}$ range, both clockwise and anticlockwise (all plotted data here are for the anticlockwise direction) at fixed temperature and field magnitude. The samples were mounted in three different orientations with respect to the cantilever platform in order to probe the torque for magnetic field relative to the crystallographic axes in three orthogonal planes, $ab$, $bc^*$ and $ac^*$. The platform rotation axis vector $\bm{u}$ is along the cross product $\bm{v}_2 \times  \bm{v}_1$, where $\bm{v}_1$($\bm{v}_2$) is the first (second) crystallograpic axis to become aligned with the field $\bm{H}$ as the angle $\theta$ of the sample platform increases in the positive sense. For example, in Fig.~\ref{F:levers}(c) $\bm{v}_1=\bm{c^*}$, $\bm{v}_2=-\bm{b}$, so $\bm{u} \parallel -\bm{b} \times \bm{c^*} \parallel -\bm{a}$. To ensure measurements of a given sample mounted in three different orientations were directly comparable, samples were mounted (in as far as was technically feasible) with a similar centre-of-mass position relative to the cantilever base.
The torque dependence on field magnitude was also measured for a selection of fixed orientations and temperatures to probe the crossing of magnetic phase boundaries as a function of field magnitude and temperature. Results were fully consistent between two different samples (S1 and S2), both almost single grain, see Appendix~\ref{A:xrays}. The final experimental phase diagrams in Sec.~\ref{S:phasediagrams} contain phase transition points from both samples.

In all plots showing the angular dependence of the torque [such as Fig.~\ref{F:bc*}(a)], $\tau_u$ is plotted after subtracting an overall angle-independent offset which was found as follows. When the torque $\tau_u$ is relatively small (in the ordered phase at low temperatures for low-to-moderate fields or in the paramagnetic phase at high temperatures) the measured piezocantilever voltage change is linear in $\tau_u$, $\Delta V=\mathcal{A}+\mathcal{B}\tau_u$. In this case, the full angular dependence $\tau_u(\theta)$ is symmetric around its vertical midpoint. Any small deviation from a perfectly balanced bridge (for which $\mathcal{A}=0$ in zero field neglecting the sample weight effect) can be accounted for by a vertical offset which ensures that the vertical midpoint of the curves $\tau_u(\theta)$ is at $\tau_u=0$. Using this method, for example for field in the $bc^*$ plane, we indeed find that all curves $\tau_u(\theta)$ at all measured temperatures in the relatively low field of 5~T in the $bc^*$ plane all cross $\tau_u=0$ at the same angles, identified with $\theta_{\pm b}$ and $\theta_{\pm c^*}$, as expected from symmetry arguments.

However, at low temperatures and very high fields when $\tau_u$ becomes substantial, the dependence $\Delta V$ vs. $\tau_u$ is still monotonic but becomes non-linear and the two bending senses of the lever are no longer symmetric, i.e. the lever bends more easily in one sense than in the other; these effects become clearly visible in the angular scans at the highest fields [see for example Fig.~\ref{F:bc*}(a)]. In these cases, vertical offsets were chosen to ensure that the torque zero crossings occur for the same angles as in the case of the same experiment in the low-field limit when the midpoint method can be applied. The absolute sign of $\tau_u$ was determined from the observed angular dependence of the very small, but detectable torque in zero applied field due to the finite sample weight. This non-magnetic torque was found to be negligible in magnitude compared to the magnetic torque for essentially all measurements reported here and was therefore not subtracted from the raw torque data.

The torque $\tau_u$ experienced by the sample is directly related to the angular derivative of the free energy $\mathcal{F}$ with respect to the rotation angle $\theta$, $\tau_u=-\frac{\partial \mathcal{F}}{\partial \theta}$ where the negative sign arises because the experimental angle $\theta$ corresponds to the rotation angle of the sample platform in a fixed field $\bm{H}$, rather than the rotation angle of the field for a fixed sample orientation. Stable equilibrium (local minimum in the free energy $\mathcal{F}$) at an angle $\theta_0$ corresponds to the torque $\tau_u$ passing through zero with a negative gradient, i.e. $\tau_u(\theta_0)=0$ with $\left. \frac{\partial \tau_u}{\partial \theta}\right|_{\theta=\theta_0} < 0$, such that in the limit of small deviations around $\theta_0$ the torque has a restoring form $\tau_u(\theta)=-\kappa(\theta-\theta_0)$ with $\kappa>0$.

\begin{figure}[tb]
\centering
\includegraphics[scale=1]{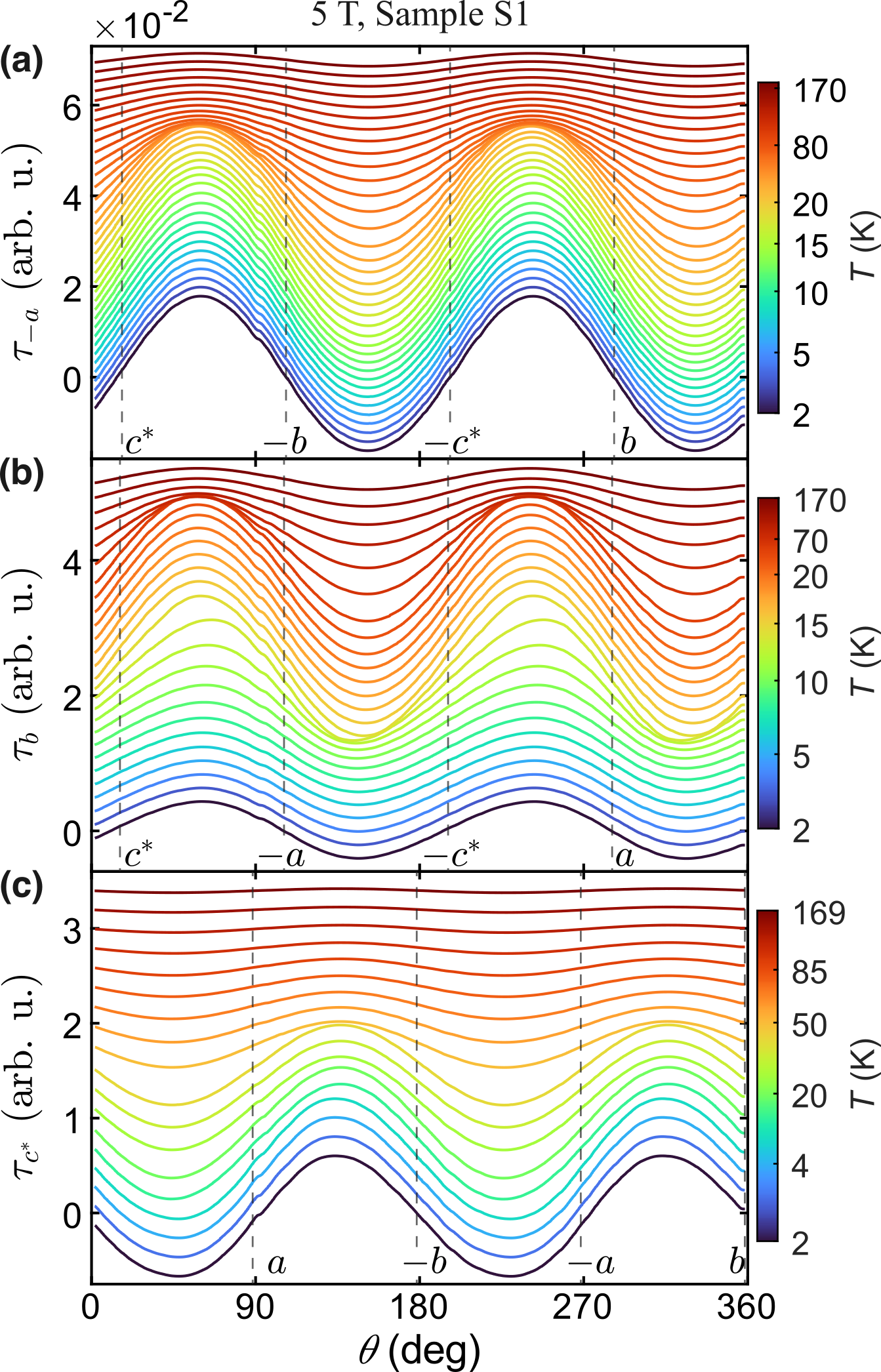}
\caption{
Angular dependence of the torque for field in the $bc^*$, $ac^*$ and $ab$ planes, as a function of temperature (color-coded) at 5~T (sample S1). In each panel, curves are offset vertically by 0.002 for clarity. Dashed vertical lines and bottom labels indicate when the applied field is oriented along reference crystallographic axes. In (c), the four lowest temperature curves are scaled by $\times0.5$ for clarity.
}
\label{F:Tdep}
\end{figure}

\section{Results}\label{S:results}
We first present the temperature-dependence of angular torque scans from which we deduce the order of the magnetic susceptibilities along three principal directions ($\chi_b>\chi_a>\chi_{c^*}$) and observe clear evidence of a sharp, single magnetic transition at $T_{\rm N}=14$~K, and then we present angle and field magnitude torque scans at low temperatures to explore the magnetic phase diagrams.

\subsection{Temperature-dependence of the torque, order of magnetic susceptibilites and determination of $T_{\rm N}$}\label{S:Tdep}

An overview of the observed angular-dependence of the torque as a function of temperature for field in the three crystallographic planes is shown in Fig.~\ref{F:Tdep}. Torque zero crossings indicate equilibrium positions, and in the limit of small field these orientations correspond to principal directions of the magnetic susceptibility tensor in that plane. In the paramagnetic phase the $2/m$ crystal lattice point group requires that $b$ is a principal axis. This is also true in the zigzag-Z phase illustrated in Fig.~\ref{F:C2m}(e), which is expected to be the ground state at low temperatures and small applied fields, the magnetic point group of this phase is also $2/m$, as explained in \cite{Johnson2015}. For field in the $bc^*$ plane in Fig.~\ref{F:Tdep}(a), over-plotting all observed torque curves relative to their vertical midpoint reveals symmetric curves with four common zero crossings (not shown), identified with orientations where $H \parallel \pm c^*$ and $\parallel \pm b$, as expected by symmetry. In this plane the mirror $m$ imposes $\tau_{-a}(\theta_{\pm c^*})=0$ and the 2-fold rotation imposes $\tau_{-a}(\theta_{\pm b})=0$. For clarity, vertically-offset curves are plotted in Fig.~\ref{F:Tdep}(a) with dashed vertical lines indicating the location of torque zero crossings. The non-zero value of $\theta_{c^*}$ is identified with the rotator motor offset angle when the lever normal is parallel to the (fixed) applied field. Similarly, common torque zero crossings for all angular torque curves are found for field in the $ab$ plane in Fig.~\ref{F:Tdep}(c), identified with $H \parallel \pm a$ and $\pm b$ orientations, as expected by symmetry. For field in the $ac^*$ plane in Fig.~\ref{F:Tdep}(b), the torque zero crossings are not symmetry constrained to occur for any particular field orientations except that each zero crossing should have a partner $180^\circ$ away, required by the 2-fold rotation axis around the plane normal, $b$-axis. However, using the midpoint offsetting method we empirically find common torque zero crossings near positions when the cantilever normal is almost along the field or orthogonal to it, which we identify with $H \parallel \pm c^*$ and $\parallel \pm a$ orientations, respectively. So we conclude that the symmetry-imposed $b$-axis together with the empirically found $a$ and $c^*$ axes are the three principal directions of the magnetic susceptibility tensor in both the paramagnetic and zigzag-Z phases. Torque zero crossings with negative gradients indicate stable equilibrium positions (largest magnetic susceptibilities). From comparing the data in the three planes we deduce that the order of the principal magnetic susceptibilities is $$\chi_b>\chi_a>\chi_{c^*}$$ over the full temperature range studied from 2 to 300~K.

\begin{figure}[tb]
\centering
\includegraphics[scale=1]{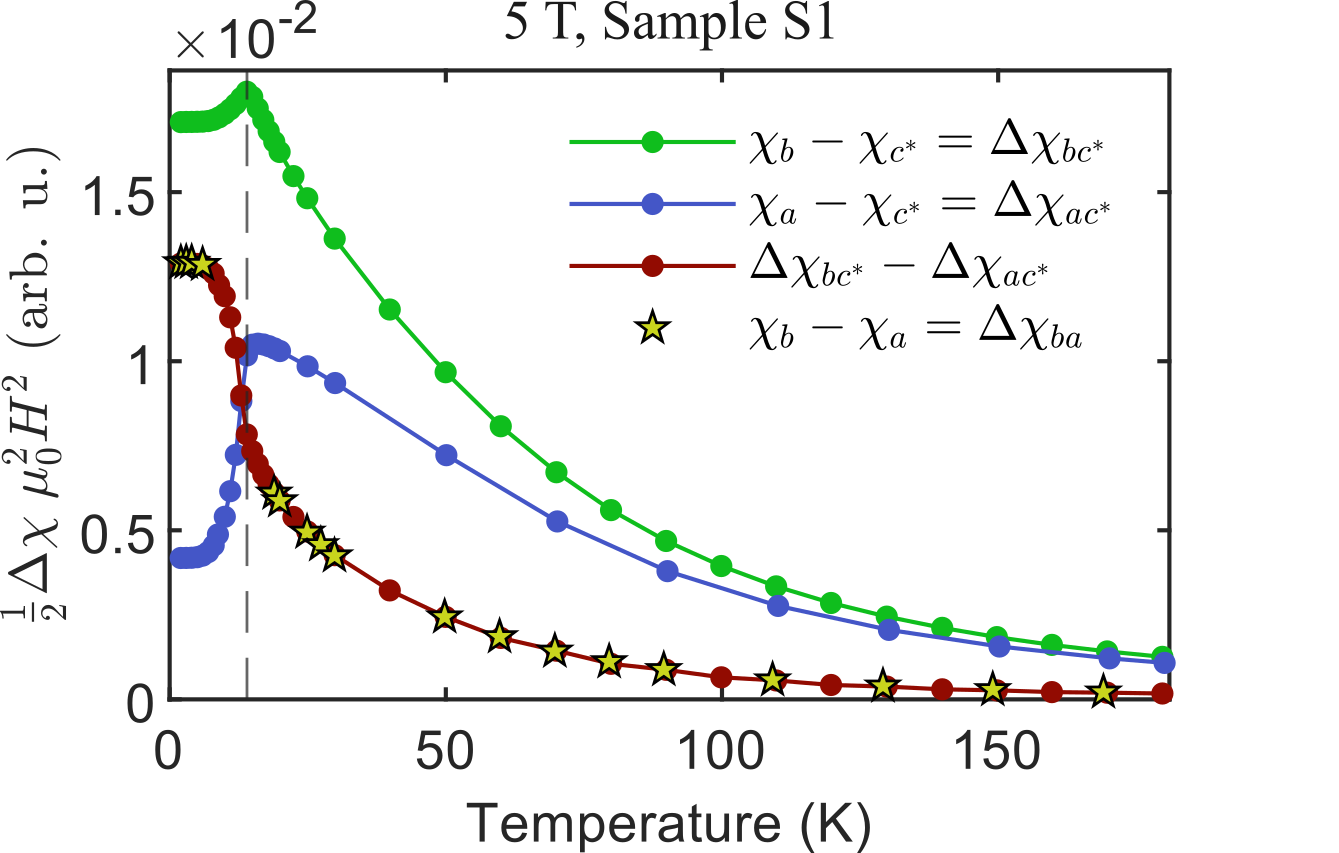}
\caption{Temperature dependence of the susceptibility anisotropy in the three crystallographic planes probed. Data points are the fitted amplitude $A_2$ of the 2-fold Fourier component of the torque angular dependence as defined in Eqs.~(\protect\ref{E:tau_a},\protect\ref{E:A2}) and described in the text. The fitted raw torque data is from Fig.~\protect\ref{F:Tdep}, collected in a field of 5~T.
The vertical dashed line indicates $T_{\rm N}$.
}
\label{F:chi}
\end{figure}

\begin{figure*}[htb]
\centering
\includegraphics[width=\textwidth]{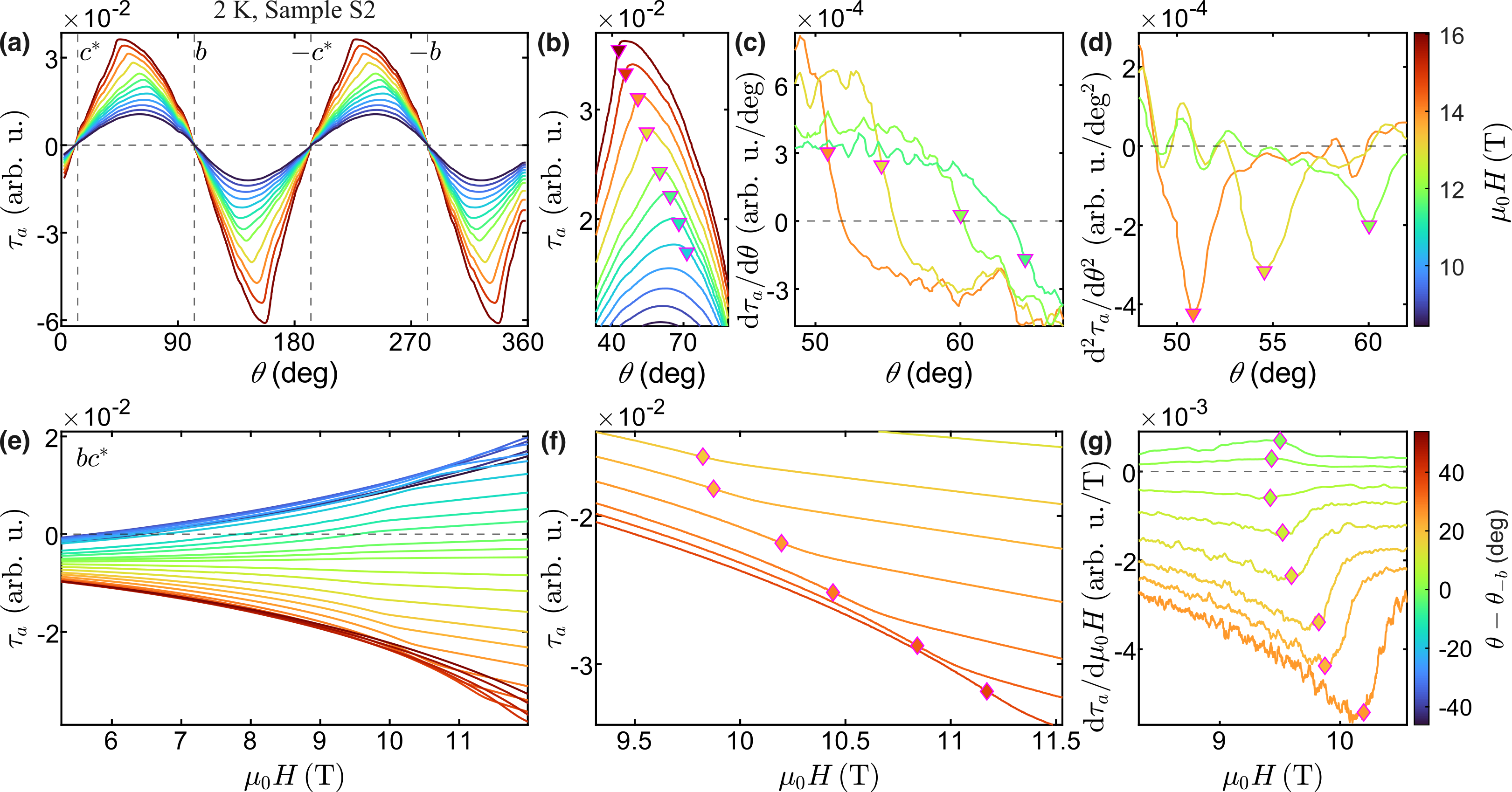}
\caption{
Torque for field in the $bc^*$ plane at 2~K (sample S2).
(a) Angular dependence of the torque as a function of field magnitude (color coded). Top labels and dashed vertical lines indicate when the applied field is orientated along reference crystallographic axes. (b) Zoomed in section of (a), where kink anomalies are indicated by down triangles. These anomalies are clearly observed as points of steepest gradient in the first derivative in (c) and as local peaks in the second derivative in (d), where only a selection of the curves are plotted for clarity. (e) Torque as a function of field magnitude for a range of angles (color coded) on both sides of the $-b$ axis. (f) Zoomed in section of (e), where diamonds indicate location of kink anomalies, clearly observed as local peaks in the first derivative in (g). All anomaly symbols are color-coded as the curves they belong to.}
\label{F:bc*}
\end{figure*}

In the paramagnetic phase and in the limit of small applied fields the torque is expected to have a two-fold periodic sinusoidal angular dependence, with the Fourier amplitude proportional to the difference in the magnetic susceptibilities along the two principal directions in that plane, i.e.
\begin{equation}
\tau_{-a}(\theta)=A_2\sin 2(\theta-\theta_{c^*})
\label{E:tau_a}
\end{equation}
for the $bc^*$ plane, with the Fourier amplitude
\begin{equation}
A_2=\frac{1}{2}(\chi_b-\chi_{c^*})\mu^2_0H^2,
\label{E:A2}
\end{equation}
where $\theta_{c^*}$ is the rotation angle when $H \parallel c^*$. The temperature-dependence of the susceptibility difference $\Delta\chi_{bc^*}=\chi_b-\chi_{c^*}$ as obtained from fits of $\tau_{-a}$ to the above functional form is plotted in Fig.~\ref{F:chi} (green symbols). A clear anomaly is observed at $T_{\rm N}=14$~K attributed to the transition from paramagnetic to the zigzag-Z antiferromagnetic order. We also note that below $T_{\rm N}$ the torque angular dependence departs only marginally from a perfect sinusoid, with a slightly steeper gradient when the field is near $c^*$ (unstable equilibrium) than near $b$ (stable equilibrium).

The obtained temperature-dependence of the susceptibility difference $\Delta \chi_{ac^*}=\chi_a-\chi_{c^*}$ in Fig.~\ref{F:chi} (blue points) shows an even stronger anomaly (steep drop) upon cooling through $T_{\rm N}$. Finally, the difference $\chi_b-\chi_a$ (brown points) deduced from subtracting the above two curves is fully consistent at the quantitative level with $\Delta\chi_{ba}$ (stars) obtained from fits to the torque data in the $ab$ plane in Fig.~\ref{F:Tdep}(c). This indicates that the torque data are fully consistent at the quantitative level between the three distinct sample orientations.

We note that $\chi_b$ being the largest susceptibility at low temperatures is naturally expected for the zigzag-Z magnetic order with moments in the $ac^*$ plane, as detected by neutron diffraction on powder samples \cite{Johnson2015}. In this case, the largest magnetization is induced by a field along the $b$-axis, normal to the antiferromagnetically-aligned moments. The observed drop in $\chi_a-\chi_{c^*}$ below $T_{\rm N}$ is consistent with moments aligned closer to the $a$ rather than the $c^*$ axis. This is expected by assuming the same in-plane magnetic order pattern as in the rhombohedral samples, where the in-plane antiferromagnetic zigzag order has Ru$^{3+}$ moments making only a small out-of-plane tilt angle $\alpha=32$–$35^{\circ}$ with the monoclinic $a$-axis \cite{cao2016,Sears2020,Kim2025}.

\subsection{Torque for field in the $bc^*$ plane}\label{S:bc*}
An overview of the angular-dependence of the torque for field in the $bc^*$ plane at the base temperature of 2~K is shown in Fig.~\ref{F:bc*}(a). At low field the torque shows a smooth two-fold, near-sinusoidal shape. However, above 9~T clear kinks become apparent [see Fig.~\ref{F:bc*}(b) (down triangles)] that move away from the $b$-axis towards the $c^*$-axis upon increasing field magnitude. Such kinks occur symmetrically on both sides of the $\pm b$ axes and are observed as local peaks in the second order derivative $\partial^2\tau_{a}/\partial\theta^2$, as illustrated in Fig.~\ref{F:bc*}(d) (down triangles). We attribute those anomalies to the transitions between the antiferromagnetic zigzag-Z order and the field-polarized paramagnetic (P) phase, as predicted by earlier theoretical calculations for $\alpha$-RuCl$_3$ \cite{Riedl2019}.

\begin{figure*}[tb]
\centering
\includegraphics[width=\textwidth]{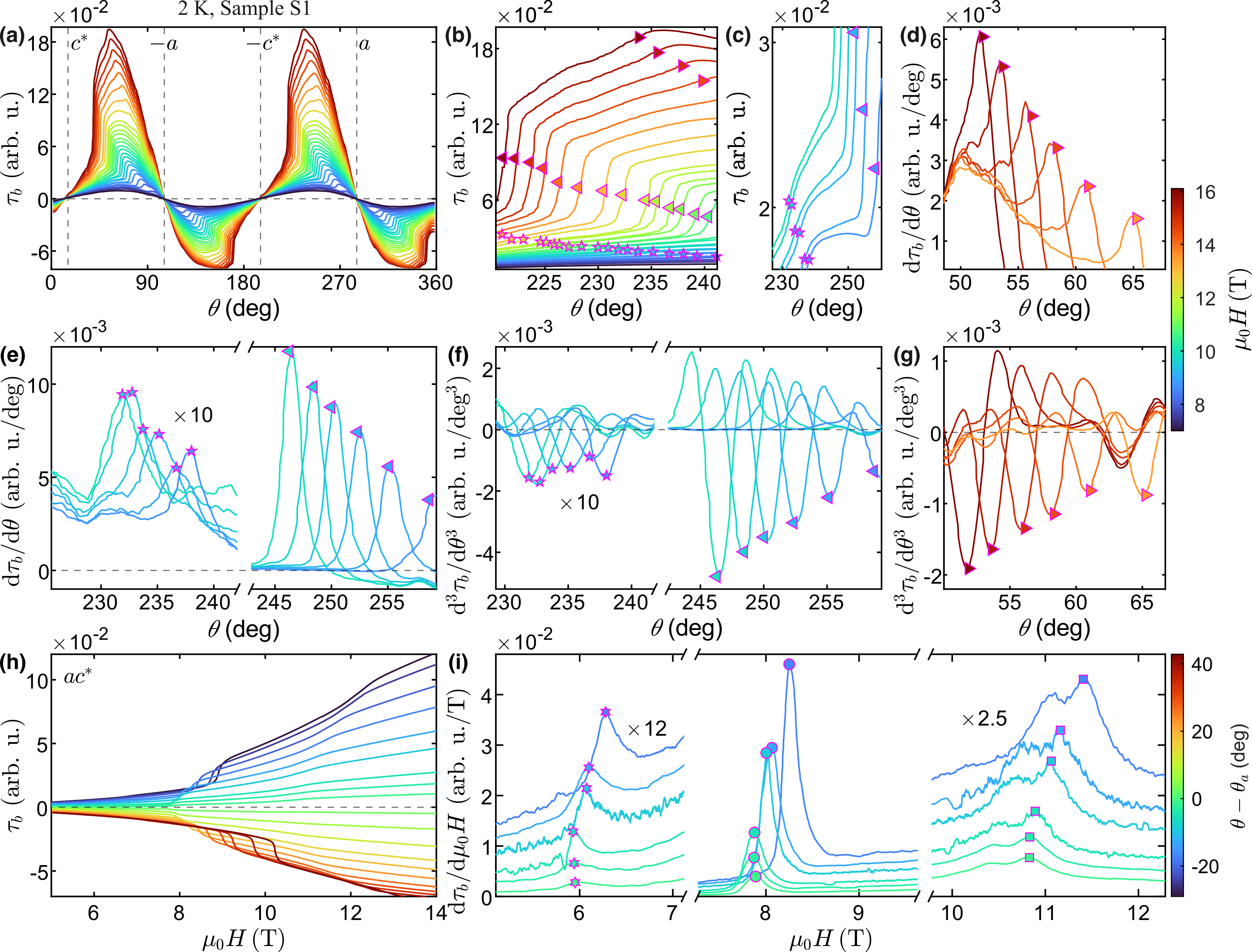}
\caption{
Torque for field in the $ac^*$ plane at 2~K (sample S1). (a) Angular dependence as a function of field magnitude (color coded). Top labels and dashed vertical lines indicate when the applied field is oriented along reference crystallographic axes. (b–c) Zoomed-in sections of (a) where the three different types of markers (stars and left/right triangles) indicate the three distinct families of anomalies described in the text. (d–e) First derivative, and (f–g) third derivative of the torque for a selection of the curves in (a) to illustrate anomalies detection. In (e–f) the curves on the left side of the horizontal axis break are scaled by the given factor for clarity. (h) Torque as a function of field magnitude for a range of angles (color coded) on both sides of the $a$-axis. (i) First derivative for a selection of the curves in (h) with left/right sections scaled by the given factors for clarity and the three distinct types of anomalies indicated by the different symbols (starts, circles and squares), as described in the text. All anomalies symbols are color-coded as the curves they belong to.
}
\label{F:ac*}
\end{figure*}

Measurements of torque as a function of field magnitude for a range of closely spaced orientations on both sides of the $-b$ axis are shown in Fig.~\ref{F:bc*}(e). The zoomed-in section in (f) illustrates the presence of kink anomalies (diamonds), appearing as local peaks in the first derivative $\partial \tau_{a}/\partial H$ shown in Fig.~\ref{F:bc*}(g). The locations of all torque anomalies for field in the $bc^*$ plane are summarized in the phase diagram in Fig.~\ref{F:phasediagrams}(c) to be discussed later in Sec.~\ref{S:phasediagrams}. There is very good agreement in the location of the anomalies detected in angular scans compared to field magnitude scans, providing an important consistency check of the criteria used above in detecting those anomalies.

Fig.~\ref{F:bc*}(a) shows that at high field when the torque magnitude is relatively large, the $\tau_a(\theta)$ curves have the negative side larger in magnitude than the positive side. This is contrary to the expectation of equal magnitude based on the $2/m$ crystal lattice point group symmetry. The observed asymmetry is attributed to the fact that at large torque magnitudes the cantilever bends easier in one sense than in the other, so negative/positive $\tau_a(\theta)$ are identified with the easy/hard senses of the lever bending. However, we find that the location of torque anomalies of both positive and negative $\tau_a$ when plotted on the phase diagram in Fig.~\ref{F:phasediagrams}(c) agree with the expected lattice point group symmetry to within the experimental accuracy, so we keep anomalies detected on both the easy and hard senses.

\subsection{Torque for field in the $ac^*$ plane}\label{S:ac*}
The angular-dependence of the torque for field in the $ac^*$ plane at 2~K is shown in Fig.~\ref{F:ac*}(a). The low-field smooth two-fold periodic, near-sinusoidal shape changes above 6~T into a much more complex shape. We identify three distinct types of anomalies, all observed to emerge from the $a$-axis and move towards $\pm c^*$ upon increasing field magnitude. The most prominent anomalies are sharp jumps in torque, indicated by left triangles in panel (b), observed as very sharp peaks in both the first derivative $\partial \tau_b/\partial \theta$ in panel (e) (right side) and the third derivative $\partial^3 \tau_b/\partial \theta^3$ in panel (f) (right side). S-shaped anomalies in panel (b) (top right, right triangles) are clearly detected as local peaks in the first derivative in panel (d) and third derivative in panel (g). Finally, there is a third family of weaker S-shape anomalies indicated by stars in the raw torque in panel (c) (bottom left), clearly detected as peaks in the first and third derivative in panels (e) and (f) (left sides). These three types of anomalies are observed on both sides of the $a$-axis and are $180^\circ$ periodic.

\begin{figure*}[tb]
\centering
\includegraphics[width=\textwidth]{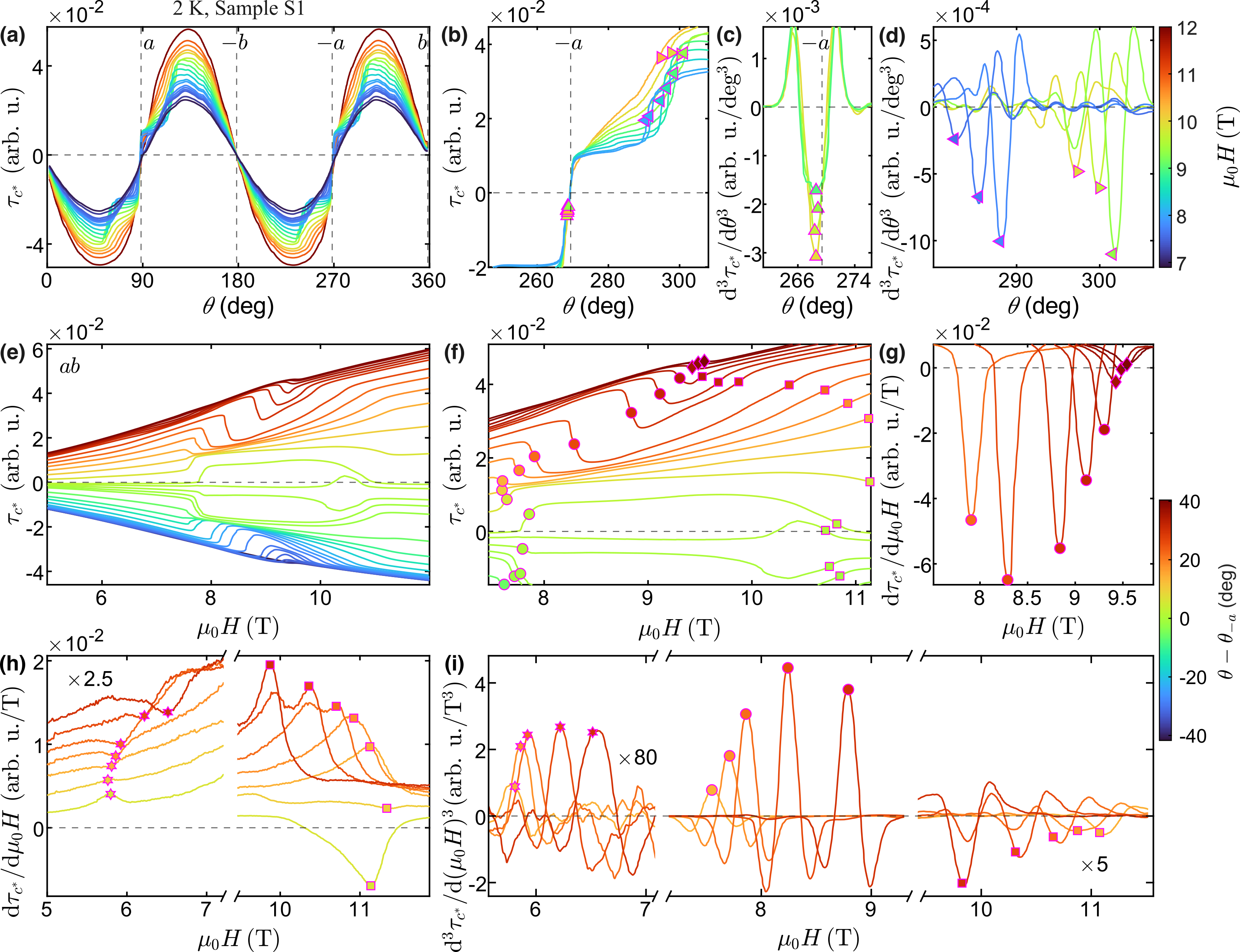}
\caption{
Torque for field in the $ab$ plane at 2~K (sample S1). (a) Angular dependence of the torque as a function of field magnitude (color coded). Top labels and dashed vertical lines indicate when the applied field is oriented along reference crystallographic axes. (b) Zoomed-in section of (a) over a restricted field range indicating jump anomalies (up triangles) and S-shaped anomalies (left and right triangles), appearing as local peaks in the third derivative in panels (c) and (d), respectively. Left/right triangles indicate anomalies that move upon increasing field away/towards the $-a$ axis indicated by the dashed vertical line in (b). (e) Torque as a function of field magnitude for a selection of angles (color coded) on both sides of the $-a$ axis. (f) Zoomed-in section of (e), with distinct symbols (circles, squares and diamonds) indicating the location of torque anomalies, clearly visible as local peaks in the first derivative in panels (g) and (h), and in the third derivative in panel (i). Horizontal axis breaks in (h) and (i) are used to separate curves that have been scaled by the given factors, for clarity. All anomalies symbols are color-coded as the curves they belong to.}
\label{F:ab}
\end{figure*}

Corresponding signatures of the above anomalies are also detected in torque scans as a function of field magnitude for a range of closely spaced orientations on both sides of the $a$-axis, with the raw torque data shown in Fig.~\ref{F:ac*}(h). Three distinct families of anomalies indicated by different symbols are detected as clear peaks in the first derivative $\partial \tau_b/\partial H$ in panel (i). The locations of all anomalies for field in the $ac^*$ plane are summarized in the phase diagram in Fig.~\ref{F:phasediagrams}(b) to be discussed later in Sec.~\ref{S:phasediagrams}. Note the good agreement in the location of the anomalies detected in the two types of scans. We note in Fig.~\ref{F:ac*} that positive/negative $\tau_b$ correspond to the easy/hard lever bending senses, with a strong asymmetry between them at the largest fields. As the torque magnitudes are somewhat larger here than in Fig.~\ref{F:bc*}(a), some corrections might be needed to the location of anomalies, however we neglect such effects and keep all anomalies in the phase diagram in Fig.~\ref{F:phasediagrams}(b).

\subsection{Torque for field in the $ab$ plane}\label{S:ab}
Finally, the angular-dependence of the torque for field in the $ab$ plane at 2~K is shown in Fig.~\ref{F:ab}(a). The low-field, smooth two-fold, near-sinusoidal shape changes above 6~T and three distinct types of anomalies can be observed. The most prominent are sharp jumps for field very close to the $\pm a$ axes, see zoomed-in region in panel (b) (up triangles), detected as sharp local peaks in the third derivative in panel (c). There are also S-shaped anomalies clearly visible in panel (b) top right corner, detected as local peaks in the third derivative in panel (d).

\begin{figure*}[tb]
\centering
\includegraphics[width=0.95\textwidth]{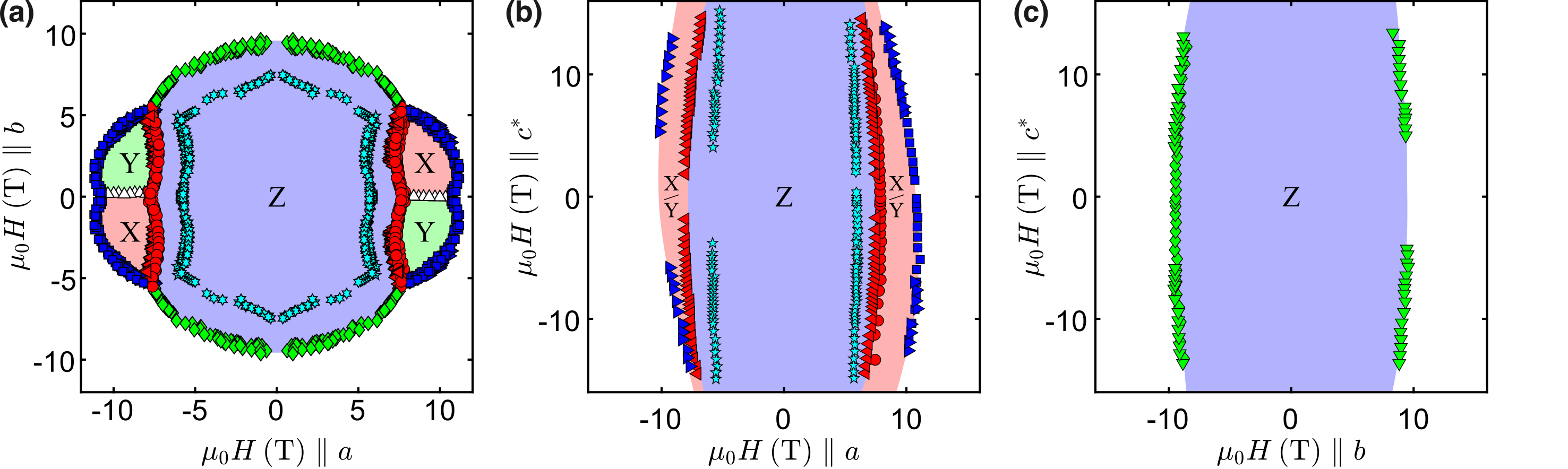}
\caption{Magnetic phase diagrams at 2~K as a function of field orientation in the three crystallographic planes explored. Symbols correspond to the locations of anomalies in torque measurements as a function of angle (circular scans) or field magnitude (radial scans starting from the origin in the above diagrams). Symbol shape is preserved from the torque scan they originate from in Figs.~\protect\ref{F:bc*}-\protect\ref{F:ab}. Color shaded areas indicate distinct regions of the phase diagram where the spontaneously-ordered antiferromagnetic spin components form the honeycomb zigzag patterns, illustrated in Fig.~\ref{F:C2m}(c,d,e), with red, green and blue corresponding to zigzag-X, -Y, and -Z, respectively. In panel (b) the X/Y label means X or Y as those two orders are degenerate for field exactly in the $ac^*$ plane. At finite field, the zigzag orders are superimposed on a uniform, field-induced magnetization. Unshaded (white) areas are the field polarized paramagnetic phase. Symbol color indicates distinct transitions, with red, blue, green and white for Z~$\leftrightarrow$~X/Y,
X/Y~$\leftrightarrow$~P, Z~$\leftrightarrow$~P and X~$\leftrightarrow$~Y, respectively. The zigzag-Z phase occupies most of the ordered phase diagram, with zigzag-X and -Y phases stable over an intermediate field range before the transition to paramagnet, when the field component in the $ab$ plane is near the $\pm a$ axes. Symbols in (b) (sample S1) and (c) (sample S2) are un-symmetrized. Symbols in (a) (samples S1 and S2) are symmetrized as described in the text. For clarity, white triangles associated with the X~$\leftrightarrow$~Y phase boundary in (a) are not symmetrized and are shown only for sample S1.}
\label{F:phasediagrams}
\end{figure*}

Torque scans as a function of field magnitude at 2~K, for a range of closely spaced orientations on both sides of the $-a$ axis are shown in Fig.~\ref{F:ab}(e). The zoomed-in section in panel (f) shows clear anomalies detected as local peaks in the first derivative in panels (g) and (h), and in the third derivative in panel (i). Anomaly locations are consistent between angle and field magnitude scans, are consistent between measurements on two distinct samples (S1 and S2), and are fully consistent with the $2/m$ lattice point group symmetry with a 2-fold rotation axis along $b$ and mirror normal to $b$. Fig.~\ref{F:phasediagrams}(a) shows all anomalies for field in the $ab$ plane, symmetrized $H_a \rightarrow -H_a$ and $H_b \rightarrow -H_b$ (for both samples). The good overlap observed between the different points provide an important consistency check of the anomalies data set.

\begin{figure}[tb]
\centering
\includegraphics[width = 0.5\textwidth]{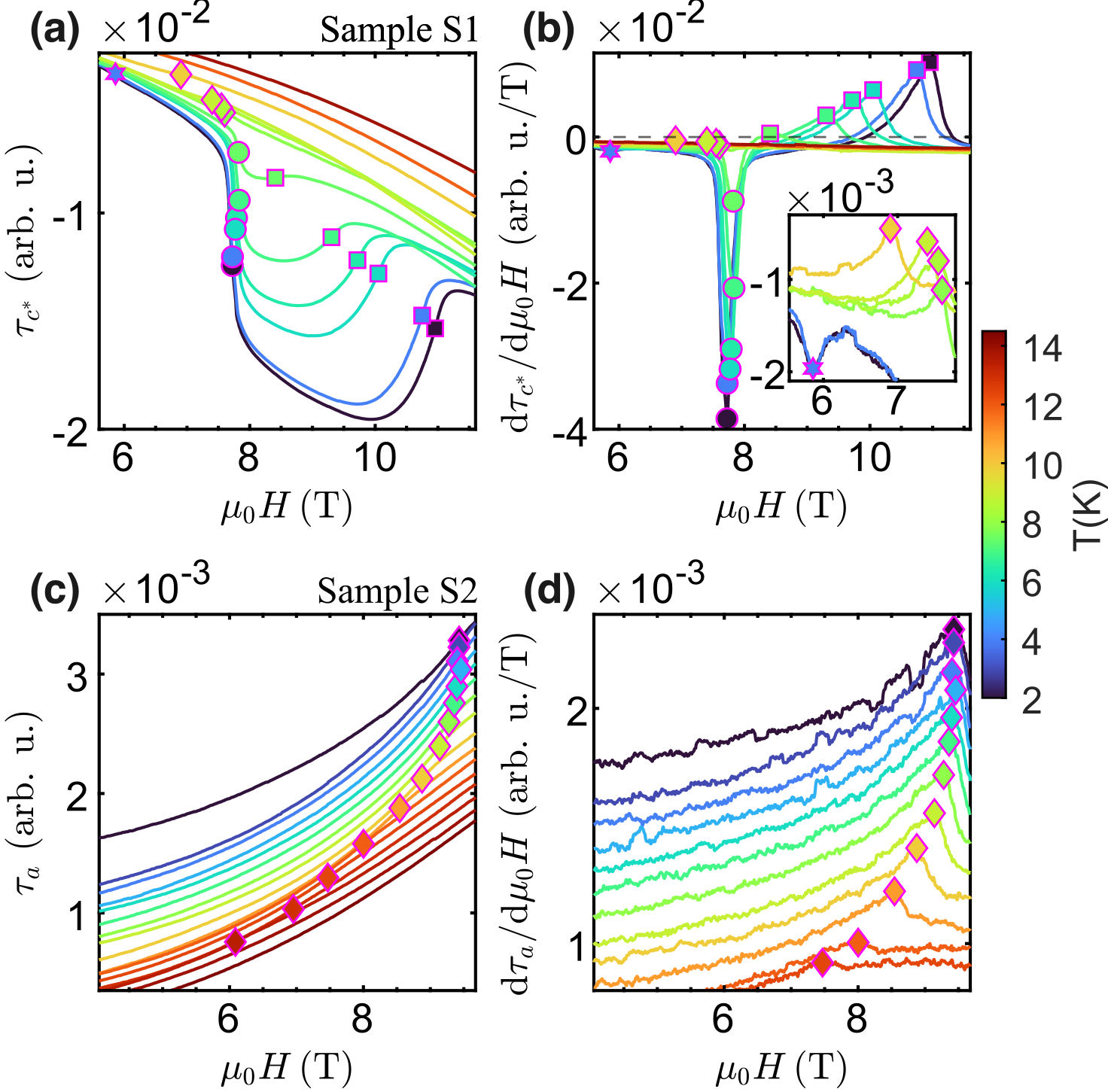}
\caption{(a) Torque as a function of field magnitude for field in the $ab$ plane near the $-a$ axis, color-coded by temperature (sample S1). (b) First derivative of the data in (a) with inset showing a zoomed-in region near 7~T. Symbols are used to make the field-temperature phase diagram in Fig.~\protect\ref{F:HT}(a). (c,d) Same as (a,b) but for field in the $bc^*$ plane near the $-b$ axis (sample S2), with the resulting phase diagram plotted in Fig.~\protect\ref{F:HT}(b). For clarity, curves in (c) and (d) are vertically offset to give a regular spacing of $0.1\times10^{-3}$ between curves (in the corresponding unit) at field values of 9~T and 7.8~T, respectively. In all panels symbols are color-coded as the curves they belong to, and indicate distinct families of anomalies (circles, squares, diamonds, and stars) plotted in the corresponding phase diagrams in Fig.~\protect\ref{F:HT}(a-b).}
\label{F:Tscans}
\end{figure}

\begin{figure}[tb]
\centering
\includegraphics[scale = 1]{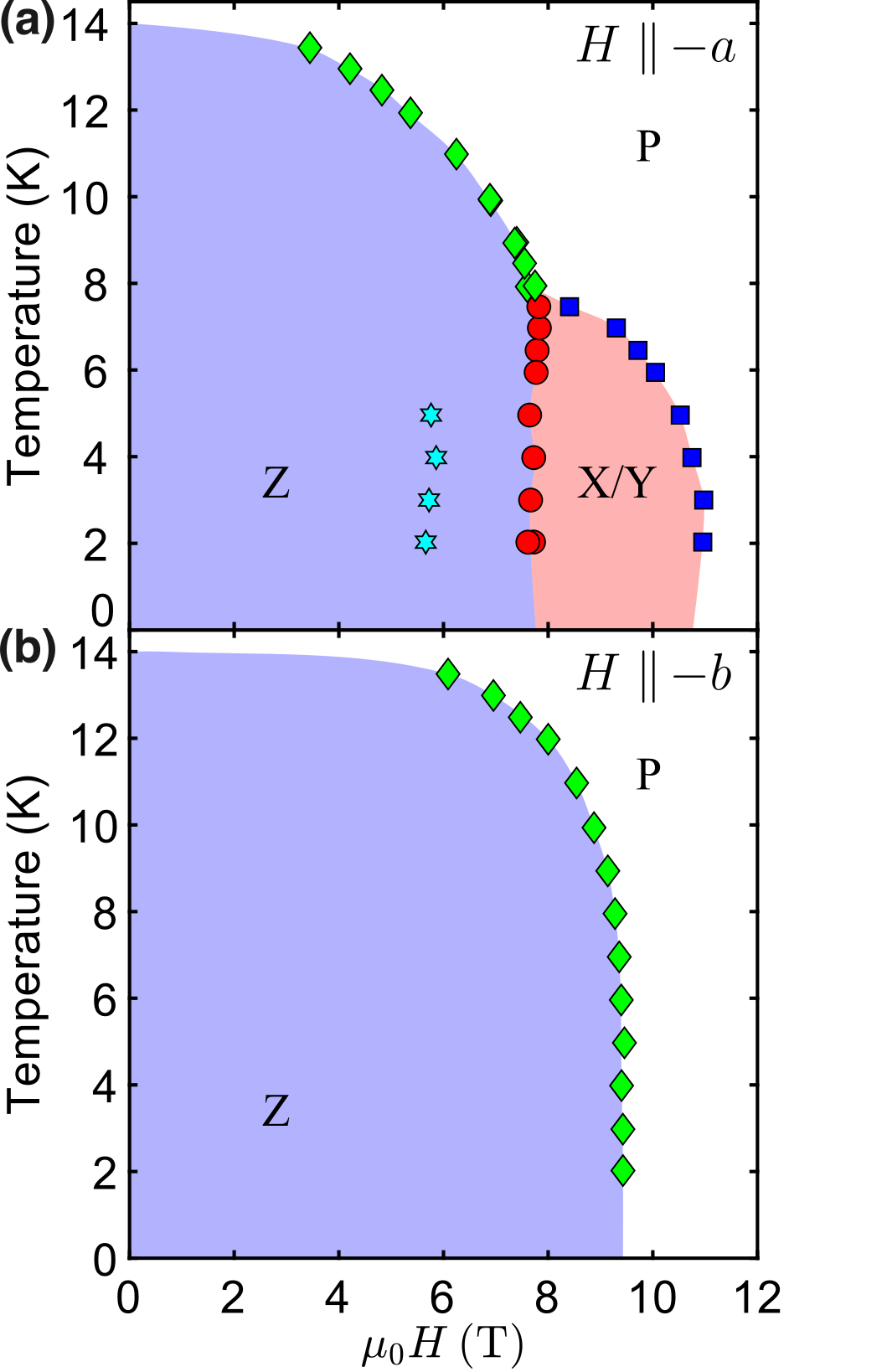}
\caption{Field-temperature phase diagram for field near the $-a$ axis, (a), and near the $-b$ axis, (b). Color shaded areas indicate distinct regions of the phase diagram where the spontaneously-ordered antiferromagnetic spin components form honeycomb zigzag patterns, illustrated in Fig.~\protect\ref{F:C2m}(c,d,e), with blue and red corresponding to zigzag-Z and zigzag-X/Y, respectively (the X/Y label means that X and Y orders are degenerate for field exactly along $\pm a$). The unshaded (white) region is paramagnetic (P). Symbols plotted in (a) and (b) are obtained from the torque anomalies identified in Fig.~\protect\ref{F:Tscans}(a-b) and (c-d), respectively. Symbol shape is preserved from the torque scan they originate from. Symbol color indicates distinct transitions, with red, blue, and green for Z~$\leftrightarrow$~X/Y, X/Y~$\leftrightarrow$~P,  and Z~$\leftrightarrow$~P, respectively. The identification of X, Y, Z, and P phases is based on comparison with mean-field phase diagrams in Sec.~\protect\ref{S:calculations}. The cyan stars in (a) correspond to very weak torque anomalies (see Fig.~\protect\ref{F:Tscans}(b) inset) that could not be detected at temperatures above 5~K.}
\label{F:HT}
\end{figure}

\subsection{Field-temperature phase diagrams}\label{S:HT}
The field-temperature phase diagrams can be determined by monitoring how the above anomalies in torque scans evolve as a function of temperature. Fig.~\ref{F:Tscans}(a) shows the temperature dependence (color coded) of torque scans as a function of field magnitude for field near the $-a$ axis. Clear anomalies are observed as peaks in the torque derivative $\partial \tau_{c^*}/\partial H$, with a systematic trend upon increasing temperature, see Fig.~\ref{F:Tscans}(b). The resulting magnetic phase diagram is plotted in Fig.~\ref{F:HT}(a). The corresponding magnetic phase diagram for a fixed field direction in the $bc^*$ plane close to the $-b$ axis is shown in Fig.~\ref{F:HT}(b). The anomalies are extracted from the raw torque scans plotted in Fig.~\ref{F:Tscans}(c), with the associated first derivative plotted in Fig.~\ref{F:Tscans}(d).

\begin{figure*}[tb]
\centering
\includegraphics[scale = 1]{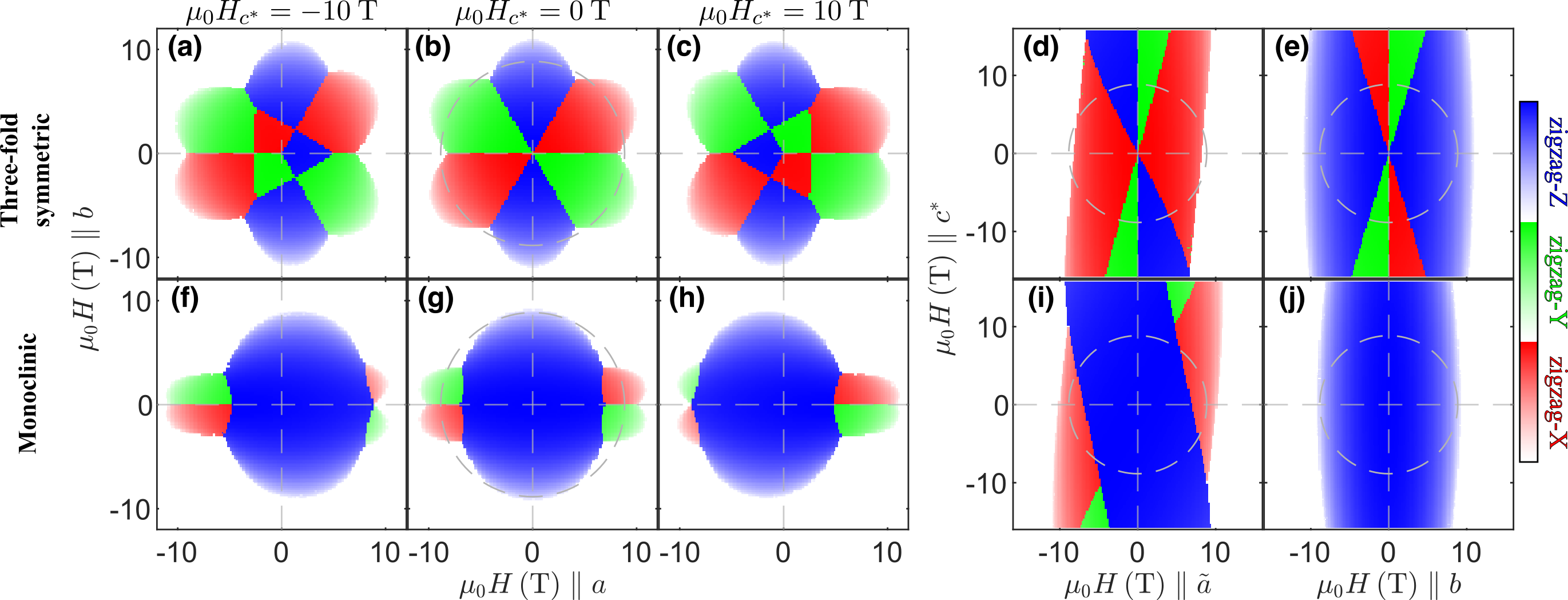}
\caption{(Top row) Mean-field magnetic phase diagrams as a function of field orientation for the three-fold symmetric Hamiltonian in Eq.~(\protect\ref{E:H3fold}). (a,b,c) are for field projected onto the $ab$ plane at $\mu_0H_{c^*}=-10$, 0 and 10~T, respectively. (d) and (e) are for field in the $\tilde{a}c^*$ and $bc^*$ planes, where the $\tilde{a}$ axis in (d) is rotated slightly away from $a$ towards $b$ to avoid exact degeneracy between zigzag-X and -Y for field exactly along $a$. Strength of red, green and blue shading indicates the order parameters in the zigzag-X, -Y, -Z phases, respectively. (Bottom row) Analogous phase diagrams for the minimal monoclinic Hamiltonian in Eq.~(\protect\ref{E:Hmono}) with $\delta_K=0.06$ and $\delta_{\Gamma}=0.3$. Dashed circles in a selection of the panels indicate the path of an angular scan at fixed field magnitude (chosen for illustration as 8.85~T to intersect phase boundaries in several planes).}
\label{F:phasediagrams_combined}
\end{figure*}

\subsection{Low-temperature magnetic phase diagrams in the $bc^*$, $ac^*$ and $ab$ planes}\label{S:phasediagrams}
The magnetic phase diagrams as a function of field in the three orthogonal planes, constructed from the above anomalies data sets are shown in Fig.~\ref{F:phasediagrams}. The color shading (red, green, blue) corresponds to the three distinct types of in-plane zigzag orders with antiparallel spin components between sites connected by X-, Y- or Z-bonds, respectively, as illustrated in Fig.~\ref{F:C2m}(c,d,e). The plotted structures show just the spontaneously ordered part of the magnetic moments, in a finite applied field this co-exists with a finite field-induced uniform polarization of the magnetic moments. We justify the above identification of the spontaneous magnetic orders in the various regions by comparing the experimental phase diagrams and raw torque data with calculations for an appropriate anisotropic spin Hamiltonian in the following sections.

\section{Minimal Spin Hamiltonian and Mean-Field Phase Diagrams}\label{S:calculations}
Here we present results of mean-field calculations of magnetic phase diagrams for candidate spin Hamiltonians with technical details provided in Appendix~\ref{A:calculations}.

\subsection{Hamiltonian for a three-fold symmetric lattice}\label{S:H3fold}
The minimal spin Hamiltonian proposed to capture the key interactions in the honeycomb layers in $\alpha$-RuCl$_3$ is the $J K \Gamma \Gamma' J_3$ model \cite{Winter_2017}. This contains nearest-neighbor couplings Heisenberg $J$, Kitaev $K$, and off-diagonal exchanges $\Gamma$ and $\Gamma'$, as well as longer range, third-neighbor Heisenberg couplings $J_3$
\begin{eqnarray}
{\mathcal H}&=&\sum_{\langle ij \rangle} J\bm{S}_i\cdot\bm{S}_j + K S_i^{\gamma}
S_j^{\gamma} + \Gamma \left(S_i^{\alpha}S_j^{\beta}+S_i^{\beta}S_j^{\alpha}\right) \nonumber \\
& & + \Gamma'\left(S_i^{\alpha} S_j^{\gamma} + S_i^{\gamma} S_j^{\alpha}+ S_i^{\beta} S_j^{\gamma} + S_i^{\gamma} S_j^{\beta}\right) \nonumber \\
& & + \sum_{\langle\langle ij\rangle\rangle} J_3 \bm{S}_i\cdot\bm{S}_j -\mu_0\mu_B \sum_i \bm{H} \cdot \mathbb{G} \cdot \bm{S}_i,
\label{E:H3fold}
\end{eqnarray}
where the first (second) sum runs over all nearest-neighbor (third-neighbor) bonds counted once with $\{\alpha,\beta,\gamma\}=\{{\sf x},{\sf y},{\sf z}\}$ for the Z-bond and permuted accordingly for the X- and Y-bonds, and the last term is the Zeeman interaction with an applied magnetic field. The above spin Hamiltonian is appropriate for a honeycomb layer where each nearest-neighbor Ru-Ru bond is inversion symmetric and a 2-fold symmetry axis, and the plane normal is a 3-fold symmetry axis. The spin axes denoted by SansSerif symbols ${\sf x}{\sf y}{\sf z}$ are along cubic Kitaev axes illustrated in Fig.~\ref{F:C2m}(b) and defined in Appendix~\ref{A:calculations}. Many distinct candidate parameter sets have been considered in the literature, for a review see Ref.~\cite{Moeller2025}. The qualitative features of the magnetic phase diagrams are common between many of the proposed parameter sets. For simplicity, we choose the minimal model proposed in \cite{Riedl2019} with $\Gamma'=0$, $(J,K,\Gamma,J_3)=(-0.5,-5.0,2.5,0.5)$~meV and the anisotropic axial $g$-tensor $\mathbb{G}$ with $(g_a,g_b,g_{c^*})=(2.3,2.3,1.3)$. For these parameters, the antiferromagnetic zigzag-X, -Y and -Z orders plotted in Fig.~\ref{F:C2m}(c,d,e) are degenerate mean-field ground states in zero applied field, related by the 3-fold honeycomb lattice symmetry. A finite in-plane magnetic field breaks the 3-fold symmetry and selects one of the three zigzag orders and the calculated magnetic phase diagram as a function of field orientation in the honeycomb plane is shown in Fig.~\ref{F:phasediagrams_combined}(b). In particular fields along the $b$-axis (parallel to the Z bonds) stabilize the zigzag-Z order (blue shaded regions), as this magnetic structure has the largest magnetic susceptibility when compared to zigzag-X or -Y: the antiferromagnetically-aligned spins are perpendicular to the applied field direction [see Fig.~\ref{F:C2m}(e)] and therefore gain the largest amount of Zeeman energy by canting towards the applied field. The zigzag-Z remains stable for in-plane field rotated by up to $\pm30^{\circ}$ away from the $b$-axis, for larger rotations that exceed those thresholds the zigzag-X (red) or -Y (green) phases are stabilized. The phase boundaries in the $(H_a,H_b)$ plane at $H_{c^*}=0$ have six-fold rotational symmetry around the origin, as expected since the Hamiltonian in Eq.~(\ref{E:H3fold}) has 3-fold rotational symmetry around the $c^*$-axis and the energies should be invariant under $\bm{H}\rightarrow -\bm{H}$ due to time-reversal symmetry. The transitions between distinct zigzag phases are 1$^{\rm st}$ order. The transitions from zigzag to paramagnetic are also in general 1$^{\rm st}$ order, and become continuous for field along the $b$-axis and six-fold rotated symmetry-equivalent positions, when the field direction is a 2-fold rotational symmetry axis of the Hamiltonian in Eq.~(\ref{E:H3fold}).

It is insightful to consider how the magnetic phase diagram changes in the case of a finite out-of-plane field component. Fig.~\ref{F:phasediagrams_combined}(a) and (c) illustrate the magnetic phase diagram as a function of in-plane field components $(H_a,H_b)$ for fixed out-of-plane field $\mu_0 H_{c^*}=\mp10$~T, respectively. A new triad of zigzag-X, -Y, -Z phases emerges for small in-plane field, i.e. for field tilted away a small amount from the $c^*$ axis, as noted in a previous theoretical study \cite{Pearce2025} of the same class of Hamiltonians. The phase diagrams in the $(H_a,H_b)$ plane at finite $H_{c^*}\neq 0$ has only 3-fold rotational symmetry around the $H_{c^*}$ axis, and the phase diagrams for $\mu_0 H_{c^*}=\mp10$~T are not mirror-symmetric, but are instead inversion-symmetric. Upon increasing $|H_{c^*}|$, the in-plane field extent of the inner triad grows and the outer zigzag regions reduce, such that at large enough $|H_{c^*}|$ only the inner triad remains, above which its extent is gradually reduced upon further increasing $|H_{c^*}|$ until the transition to the field polarized paramagnetic phase (not shown). The magnetic phase diagrams for (low) field in the $ac^*$ and $bc^*$ planes are illustrated in Fig.~\ref{F:phasediagrams_combined}(d) and (e), respectively, note the change in the character of the zigzag phase upon rotating field in either plane.

\subsection{Hamiltonian for a monoclinic lattice}
For a monoclinic lattice, the Z bonds in Fig.~\ref{F:C2m}(b) are symmetry distinct from the X and Y bonds, which remain symmetry-related by a 2-fold rotation around the Z bond that bisects them. We find empirically, that a minimal spin Hamiltonian which can account qualitatively [and to a large extent also quantitatively] for the experimental phase diagram requires monoclinic anisotropy only in the $K$ and $\Gamma$ exchanges. The other types of exchanges in the spin Hamiltonian in Eq.~(\ref{E:H3fold}) are assumed to still be 3-fold symmetric. Following the theoretical band-structure studies in Ref.~\cite{Cen2025}, we introduce dimensionless parameters $\delta_{K},\delta_{\Gamma}$ to characterize the monoclinic anisotropy of the $K$ and $\Gamma$ nearest-neighbour exchanges on the X-, Y- and Z-bonds as
\begin{eqnarray}
K_{\rm Z}=K &,& K_{\rm X,Y}=K(1+2\delta_K),\nonumber \\
\Gamma_{\rm Z}=\Gamma &,&\Gamma_{\rm X,Y}=\Gamma(1-\delta_{\Gamma}).
\label{E:Hmono}
\end{eqnarray}
Magnetic phase diagrams for this minimal monoclinic model are plotted in Fig.~\ref{F:phasediagrams_combined}(bottom row). For monoclinic bond anisotropy parameters $\delta_{K},\delta_{\Gamma}>0$, as expected based on band-structure calculations when Z is the shortest bond \cite{Cen2025}, the zigzag-Z phase becomes the lowest energy state in zero field, and intermediate in-plane fields close to the $a$-axis can stabilize the zigzag-X or -Y phases, as illustrated in Fig.~\ref{F:phasediagrams_combined}(g).

\section{Comparison of Torque Data with the Minimal Monoclinic  Hamiltonian}\label{S:comparison}

The key features of the experimental phase diagrams in all three crystallographic planes probed are well captured by the mean-field calculations for the minimal monoclinic Hamiltonian. In particular there is a single ordered phase in the $bc^*$ plane attributed to zigzag-Z order with the boundary to paramagnetic only weakly dependent on $H_{c^*}$ for the field range probed, compare Fig.~\ref{F:phasediagrams}(c) with Fig.~\ref{F:phasediagrams_combined}(j). In contrast, in the $ab$ plane the phase boundaries are very strongly-field dependent with the paramagnetic phase reached within the probed field range for all orientations. Furthermore, there are two additional distinct phases observed at intermediate field near the $a$-axis attributed to field-stabilized zigzag-X and -Y orders, with a phase boundary between them when $H_b$ changes sign, compare Fig.~\ref{F:phasediagrams}(a) with Fig.~\ref{F:phasediagrams_combined}(g). The phase diagram in the $ac^*$ plane (see Fig.~\ref{F:phasediagrams}(b)) shows that the intermediate field ordered phase for field near the $a$ axis persists for finite $H_{c^*}$ fields, in qualitative agreement with the calculation in Fig.~\ref{F:phasediagrams_combined}(i) showing persistency of the red region at finite $H_{c^*}$, although with a faster rate of change in the field boundaries with $H_{c^*}$ than seen experimentally.

The values for $\delta_{K}$ and $\delta_{\Gamma}$ were adjusted to reproduce as close as possible the experimental transition fields for field along $a$ : Z$\leftrightarrow$X/Y$\leftrightarrow$ P (7.8(1) and 10.7(1)~T), for field along $b$ : Z $\leftrightarrow$ P (9.5(3)~T) and the angular extent of the X/Y phases in the $ab$ plane away from the $a$-axis (34(3)$^{\circ}$), with the calculated values being 7, 10.2, 9~T and 28$^\circ$, respectively; the rest of the Hamiltonian parameters were unchanged compared to Ref.~\cite{Riedl2019}.

\begin{figure*}[tb]
\centering
\includegraphics[scale = 1]{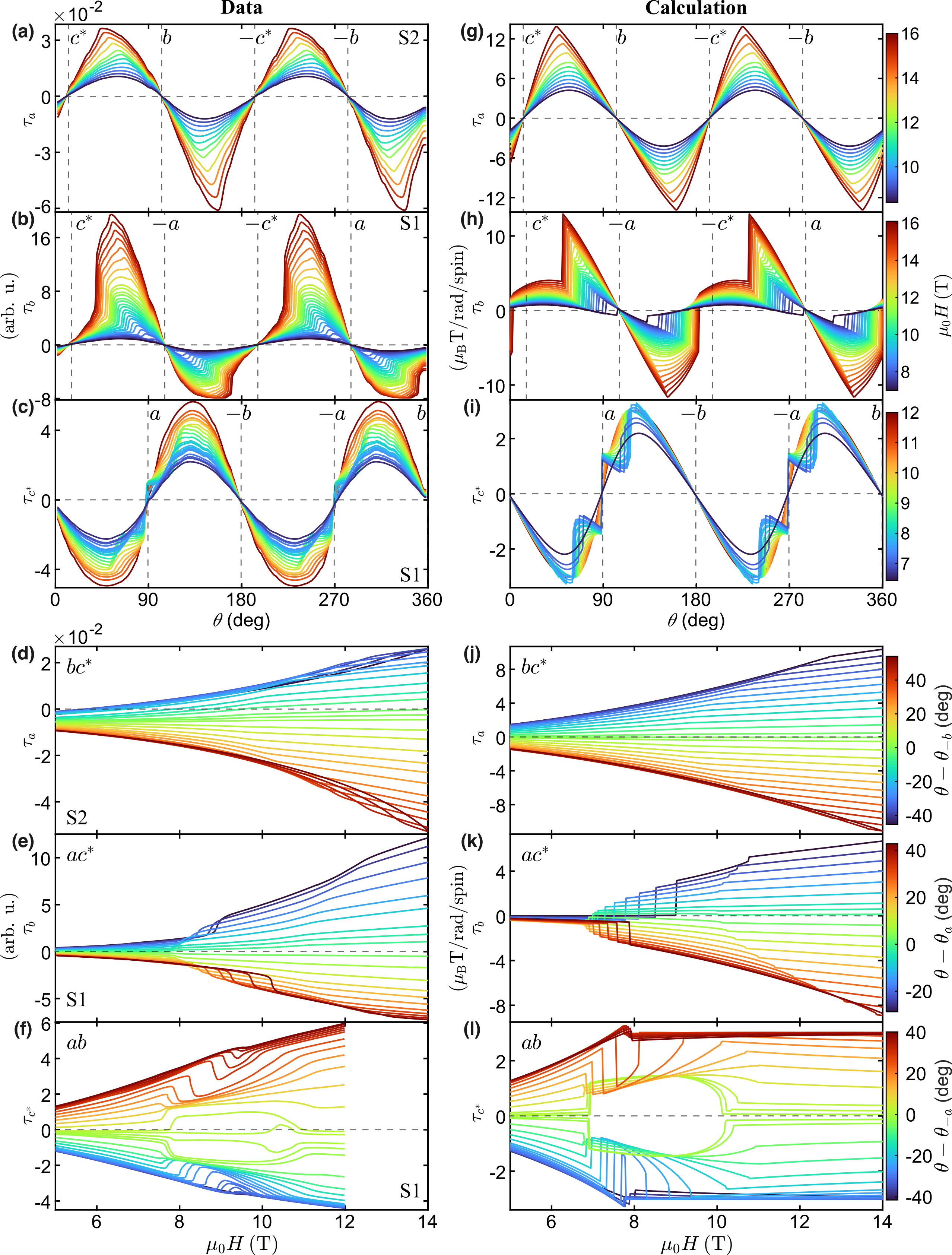}
\caption{Angular dependence of the torque in three crystallographic planes (top three rows, curves color-coded by field magnitude) and field-dependence of the torque for a range of angles in those planes (bottom three rows, curves color-coded by angle), (left) experimental data and (right) corresponding calculations for the same angles and fields for the minimal monoclinic Hamiltonian described in the text. Torque is in absolute units in the calculation panels and arbitrary units in the data panels.}\label{F:torque_calc}
\end{figure*}

It is insightful to compare not just the shape of the phase diagrams, but also the detailed torque dependence on field orientation and magnitude, and the specific signature of anomalies in the torque associated with crossing phase transition boundaries. Below we describe this comparison for each crystallographic plane in turn. Fig.~\ref{F:torque_calc}(a) data and (g) calculation show the torque angular-dependence in the $bc^*$ plane with the calculated phase diagram in Fig.~\ref{F:phasediagrams_combined}(j), where the dashed circle indicates the path of an angular scan at fixed field magnitude of 8.85~T. The torque angular dependence changes from sinusoidal at low fields when the measurement path is contained entirely inside the zigzag-Z phase, to a saw-tooth shape with clear, acute kink anomalies associated with crossing the phase boundary Z $\leftrightarrow$ P, with very clear similarities between data and calculation. The torque as a function of field magnitude in this plane, i.e. radial scans in Fig.~\ref{F:phasediagrams_combined}(j), for a range of fixed orientations is shown in Fig.~\ref{F:torque_calc}(d) data and (j) calculation. Kink anomalies separating regions with high $\to$ low slope associated with crossing the Z $\to$ P phase transition are clearly visible in both panels in the bottom/top-right corners.

The torque behavior is considerably richer for field in the $ab$ plane with the calculated phase diagram in Fig.~\ref{F:phasediagrams_combined}(g) where circular and radial scans cross multiple phase boundaries. The torque angular dependence in Fig.~\ref{F:torque_calc}(c) data and (i) calculation observe a simple sinusoidal shape at low field inside the zigzag-Z region. At higher field magnitude clear sharp jumps appear in torque when the field orientation crosses the $\pm a$ axes associated with crossing the X $\leftrightarrow$ Y phase boundary. Additional anomalies located symmetrically on either side of crossing the $\pm a$ axes are also observed, associated with Z $\leftrightarrow$ X/Y or X/Y $\leftrightarrow$ P phase transitions, with good similarity between data and calculation. Torque scans as a function of field magnitude in Fig.~\ref{F:torque_calc}(f) and (l) show a very complex evolution upon increasing angle (color coded). Starting at orientations far away from the $-a$ axis, lower traces (blue shading) observe a single anomaly associated with the direct Z $\to$ P transition, this then splits into two anomalies associated with the Z $\to$ X $\to$ P transitions with the two anomalies moving in opposite directions as the field approaches the $-a$ axis, with this behavior mirrored when the field approaches the $-a$ axis from the opposite direction when the intermediate-field phase changes character from X to Y. The distinct shape of the anomalies (near-jumps at the Z $\to$ X/Y transitions, change in slope at the X/Y $\to$ P transitions, and S-shape kinks at the Z $\to$ P transitions), as well as every sign change in those anomalies upon increasing angle are all well captured by the calculation.

Fig.~\ref{F:torque_calc}(f) shows that the torque curves are sign-reflected as the field crosses the $-a$ axis
, as expected since in that case the field orientation passes through the mirror plane of the $2/m$ point group of the crystal structure. A notable feature is that upon crossing two special angles displaced symmetrically away from the $-a$ axis, $\theta_{\pm}=\theta_{-a}\pm 10(2)^{\circ}$, the torque anomalies at both the lower and upper transitions fields appear to change sign simultaneously, {\em without} the overall torque changing sign. This non-trivial effect is captured by the calculation in Fig.~\ref{F:torque_calc}(l) near $\theta_{\pm,{\rm calc}}=\theta_{-a} \pm 8.6(1)^{\circ}$ and this can be understood as follows. The measured torque is $\tau_u=m_\eta\mu_0H$, where $m_\eta=\mathbf{m}\cdot\mathbf{\eta}$ is the component of the magnetic moment vector $\mathbf{m}$ along the direction $\mathbf{\eta}=\mathbf{\hat{H}}\times\mathbf{\hat{u}}$ that is perpendicular to {\em both} the applied field direction $\mathbf{\hat{H}}$ and the cantilever axis $\mathbf{\hat{u}}$. Focusing on the lower-field, Z $\to$ X transition for field orientation angles near $\theta_{-}$, the component $m_\eta$ decreases in magnitude upon crossing the transition for $\theta < \theta_{-}$, but increases for $\theta > \theta_{-}$, which explains why the torque anomaly at this transition changes from an upturn to a downturn as $\theta$ increases through $\theta_{-}$. Similarly, the sign of all the other torque anomalies can be related to the behavior of the magnetization component $m_\eta$.

The torque also shows multiple anomalies in the $ac^*$ plane, with the calculated phase diagram in Fig.~\ref{F:phasediagrams_combined}(i). The torque angular dependence in Fig.~\ref{F:torque_calc}(b) data and (h) calculation observe clear jumps and acute kinks associated with Z $\leftrightarrow$ X/Y $\leftrightarrow$ P phase transitions, respectively, with very good qualitative similarity between data and calculation in the form of the anomalies and trends upon increasing field magnitude, although the precise location of anomalies are only qualitatively reproduced. Torque scans as a function of field magnitude are shown in  Fig.~\ref{F:torque_calc}(e) data and (k) calculation. The presence of distinct lower and higher field, jump- and kink-like anomalies associated with the Z $\to$ X/Y $\to$ P transitions, respectively, is well reproduced, but the precise location is only approximately reproduced.

In the experimental phase diagrams in Fig.~\ref{F:phasediagrams}(a-b), inside the blue shaded central region attributed to zigzag-Z order, there are additional torque anomalies observed experimentally (cyan stars), which are not explained by the current minimal model. Those points correspond to very weak anomalies, of magnitude in the torque derivatives of order at most a few percent compared to the higher-field principal anomalies (indicated by red, green and blue symbols) and their possible origin is briefly addressed in the following Sec.~\ref{S:R-3}.

From the above comparison we conclude that the minimal monoclinic Hamiltonian provides a conceptual framework to understand the origin of the different phases in the experimental magnetic phase diagrams, and how transitions between those distinct phase lead to characteristic anomalies in torque scans as a function of field magnitude and angular orientation. We attribute the remaining quantitative discrepancies between the torque data and calculation for the minimal model to further refinements of the multi-parameter exchange Hamiltonian, including effects of weak but finite inter-layer couplings, or beyond-mean-field effects, which we expect will not change the qualitative behavior, but could further improve the quantitative agreement with the experiment.

\section{Phase Diagram of Monoclinic vs. Rhombohedral samples}
\label{S:R-3}

\begin{figure}[tb]
\centering
\includegraphics[scale = 1.2]{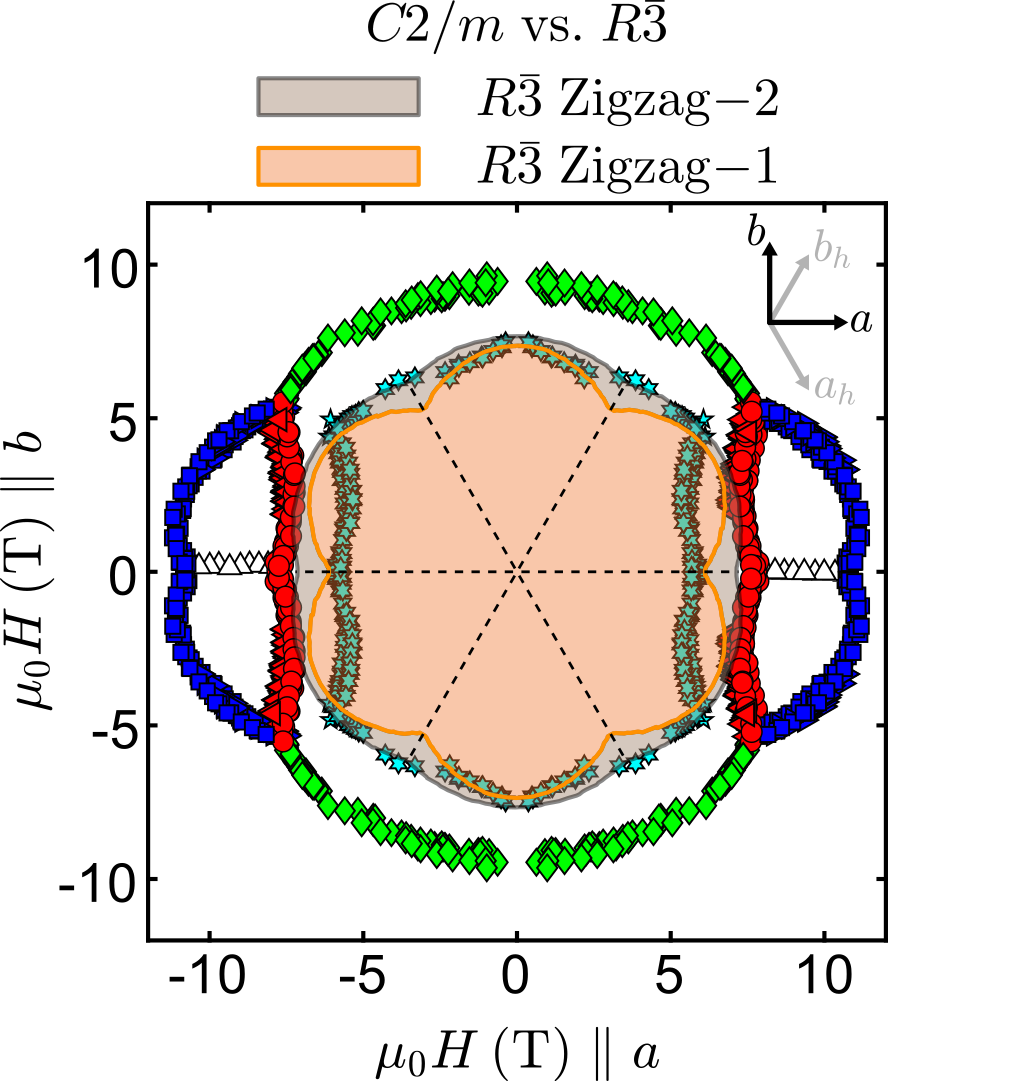}
\caption{Comparison of the $ab$ plane magnetic phase diagram in the present monoclinic ($C2/m$) samples (points from Fig.~\protect\ref{F:phasediagrams}(a)) with that of rhombohedral ($R\bar{3}$) samples (shaded regions bounded by solid lines from \protect\cite{Balz2021}). The radial dashed lines every 60$^\circ$ emphasize the relation with the generic phase diagram in Fig.~\protect\ref{F:phasediagrams_combined}(b) expected for a 3-fold symmetric Hamiltonian, with the 6 lobes corresponding to alternating zigzag-X, -Y and -Z in-plane orders. Zigzag-1(2) are collective names for all those in-plane orders, with 1(2) indicating distinct layer stacking periodicity. Top right inset shows the orientation of the hexagonal $R\bar{3}$ (grey) relative to the monoclinic (black) unit cell basis vectors.
}
\label{F:Literature_PhaseDiagram}
\end{figure}

It is insightful to compare the low-temperature experimental magnetic phase diagram of the present monoclinic ($C2/m$) samples with that of the more extensively studied samples with a low-temperature rhombohedral ($R\bar{3}$) crystal structure. A direct comparison for field in the honeycomb $ab$ plane is shown in Fig.~\ref{F:Literature_PhaseDiagram}, where symbols are as in Fig.~\ref{F:phasediagrams}(a) and top left inset shows the relative orientation of hexagonal ($R\bar{3}$) unit cell basis vectors (gray arrows and labels with subscript $_h$) relative to the monoclinic axes: $\mathbf{a}_h+\mathbf{b}_h \parallel \mathbf{a}$ and $-\mathbf{a}_h+\mathbf{b}_h \parallel \mathbf{b}$ \cite{Kim2025}. Solid lines show phase boundaries for the rhombohedral samples, obtained from location of anomalies in susceptibility measurements as a function of field magnitude and orientation \cite{Balz2021}, with the different color shading indicating zigzag orders with different periodicity in the layer stacking direction. A very noticeable difference is that all transition fields to the high-field paramagnetic phase are much larger in magnitude in the monoclinic case (10.7~T~$\parallel a$ and 9.5~T~$\parallel b$) compared to the rhombohedral case (7.1~T~$\parallel a$ and 7.7~T~$\parallel b$), where $a$ and $b$ are the monoclinic axes. In the rhombohedral case the phase boundaries have 6-fold rotational symmetry with 6 lobes (separated by dashed radial lines in Fig.~\ref{F:Literature_PhaseDiagram}) corresponding to symmetry-equivalent zigzag-X,-Y,-Z phases, whereas in the monoclinic case the reduced symmetry favors the zigzag-Z phase in the central part of the phase diagram with the zigzag-X and -Y phases stable only over a narrow, intermediate field range for fields near the $a$-axis.

Finally, we briefly address the possible origin of the very weak torque anomalies observed in the present samples in the range 5.5-7.9~T indicated by cyan stars in Fig.~\ref{F:phasediagrams}(a). Those anomalies are very weak compared the principal torque anomalies, so we attribute them to very weak, subleading effects. As the fields where those anomalies occur are well below any of the transition fields associated with principal torque anomalies observed in {\em any} field orientation in the phase diagrams in Fig.~\ref{F:phasediagrams}(a-c), those weak anomalies cannot be attributed to originating from other misaligned minority monoclinic twins in the samples. So even though we do show in Appendix~\ref{A:xrays} that the measured samples contained additional monoclinic twins of order few percent weight, the weak torque anomalies cannot be attributed to those twins, they must have a different origin.

It is notable from Fig.~\ref{F:Literature_PhaseDiagram} that the unexplained weak anomalies do overlap well with the experimentally-determined phase boundaries of rhombohedral samples for a substantial angular range around the $\pm b$ axes, so it would be tempting to attribute those unexplained very weak torque anomalies as originating from a few percent minority rhombohedral phase in our samples. However, this could not consistently explain all the weak anomalies observed, in particular those near the $a$-axis and a wide angular range around it, where the weak anomalies occur at fields well below where the rhombohedral phase boundaries would have been. Given all the above we propose that those weak anomalies could be due to magnetic transitions that occur in a minority phase (of at most a few percent weight) of a type of structural layer stacking distinct from the well-characterized monoclinic $C2/m$ or rhombohedral $R\bar{3}$, or could be due to very subtle changes in the magnetic structure of the monoclinic phase associated with the inter-layer magnetic periodicity. Since those very weak anomalies could not be detected in all samples probed (they were not detected in the two-twin sample S3 discussed in Appendix~\ref{A:twins}), the former scenario is more likely.

\section{Conclusions}\label{S:conclusions}

We have used high-sensitivity angle-dependent torque magnetometry to determine the magnetic phase diagrams of small, high-quality single crystals of $\alpha$-RuCl$_3$ that retain the monoclinic $C2/m$ structure to low temperatures. In contrast to the more commonly studied rhombohedral samples, these crystals display a single sharp magnetic transition at $T_{\rm N}=14$~K. The temperature-dependent torque establishes the ordering of the principal magnetic susceptibilities, $\chi_b>\chi_a>\chi_{c^*}$, and is consistent with zigzag-Z order as the low-field magnetic ground state. By combining near-continuous angular scans with measurements as a function of field magnitude, and identifying anomalies in both the raw torque and its higher order derivatives, we have constructed complete magnetic phase diagrams up to 16~T for fields rotated in the three orthogonal $ab$, $ac^*$ and $bc^*$ crystallographic planes.

The resulting phase diagrams reveal a pronounced dependence on field orientation and reflect the reduced symmetry of the monoclinic lattice. Zigzag-Z order occupies most of the low-field ordered region. For fields close to the $a$ axis, additional intermediate-field phases are stabilized before the transition to the field-polarized paramagnetic phase and through comparison with mean-field calculations we identify these phases as zigzag-X and zigzag-Y. A minimal extension of the $JK\Gamma J_3$ Hamiltonian, incorporating monoclinic bond anisotropy in the Kitaev and $\Gamma$ exchanges, accounts for the key features of all three experimental phase diagrams. Importantly, the model reproduces not only the overall morphology and characteristic field scales, but also the distinct jumps, kinks, and complex sequence of sign changes observed when the angular- and field-dependent torque scans cross different phase boundaries. The detailed torque response therefore provides a substantially more stringent constraint on the spin Hamiltonian and nature of the magnetic orders than just the phase diagrams alone.

The monoclinic phase diagrams differ markedly from those of rhombohedral $\alpha$-RuCl$_3$, displaying reduced rotational symmetry, substantially higher fields required for the suppression of the spontaneous zigzag order, and field-induced zigzag-X and -Y phases confined to orientations near the $a$ axis. These results demonstrate that relatively modest structural changes and the associated bond dependence of the exchange interactions can strongly reorganize the magnetic phase diagram of $\alpha$-RuCl$_3$. They also establish a quantitative framework for understanding the magnetism of small, thin, strained, or otherwise structurally constrained samples that remain monoclinic at low temperature, and emphasize the importance of structural details for the understanding of the complex magnetic field behavior of this strongly spin-orbit coupled honeycomb magnet.

\begin{acknowledgments}
We acknowledge insightful discussions with Hae-Young Kee and we thank Roser Valent\'{i} for previous collaboration on related topics. This research was partially supported by the European Research Council under the European Union’s Horizon 2020 research and innovation programme Grant Agreement Number 788814 (EQFT). D.A. acknowledges support from a doctoral studentship from the above ERC grant and from Lincoln College, Oxford. J.S.P  acknowledges financial support from the EPSRC studentship EP/W524311/1, scholarship funding from the Department of Physics, via OxPEG, and the Leathersellers' Scholarship from St. Catherine's College, Oxford. A.I.C. acknowledges an EPSRC Open Fellowship (UKRI3274) and the EPSRC Centre for Doctoral Training in Superconductivity (EP/Y035453/1).
The work at Princeton University was supported by the Gordon and Betty Moore Foundation grant number GMBF-9066. Crystal structure figures were made using \textsc{vesta} \cite{Momma2011}.
\end{acknowledgments}

\newpage
\appendix

\section{X-ray diffraction studies}\label{A:xrays}
\begin{figure}[b]
\centering
\includegraphics{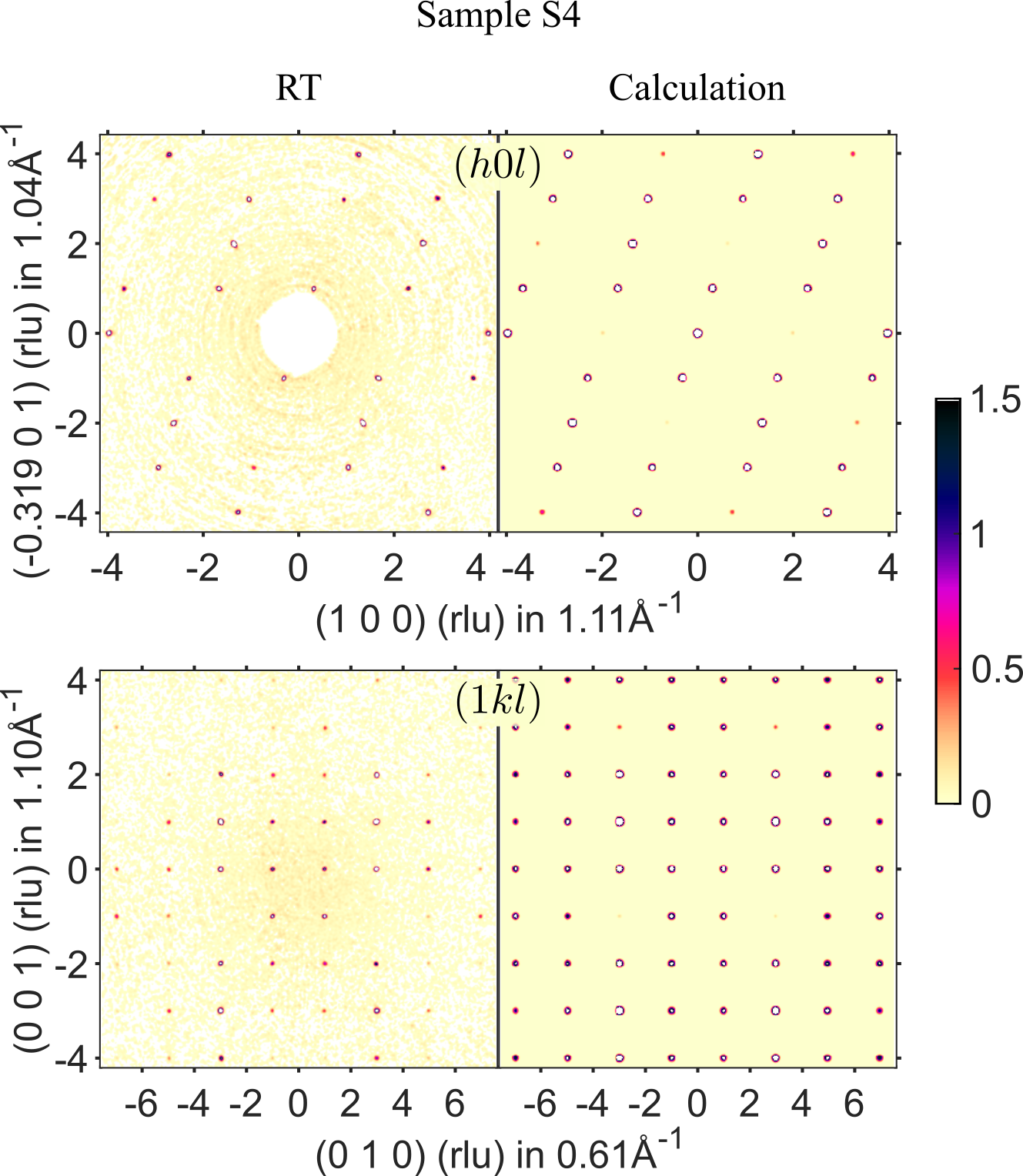}
\caption{Observed (left) vs. calculated (right) X-ray diffraction patterns at room temperature for the reference un-twinned sample S4. Structural parameters are from Table~\protect\ref{T:XRD} and color shows diffraction intensity on a log$_{10}$ scale.}
\label{F:Diff}
\end{figure}

\begin{figure}[b]
\centering
\includegraphics{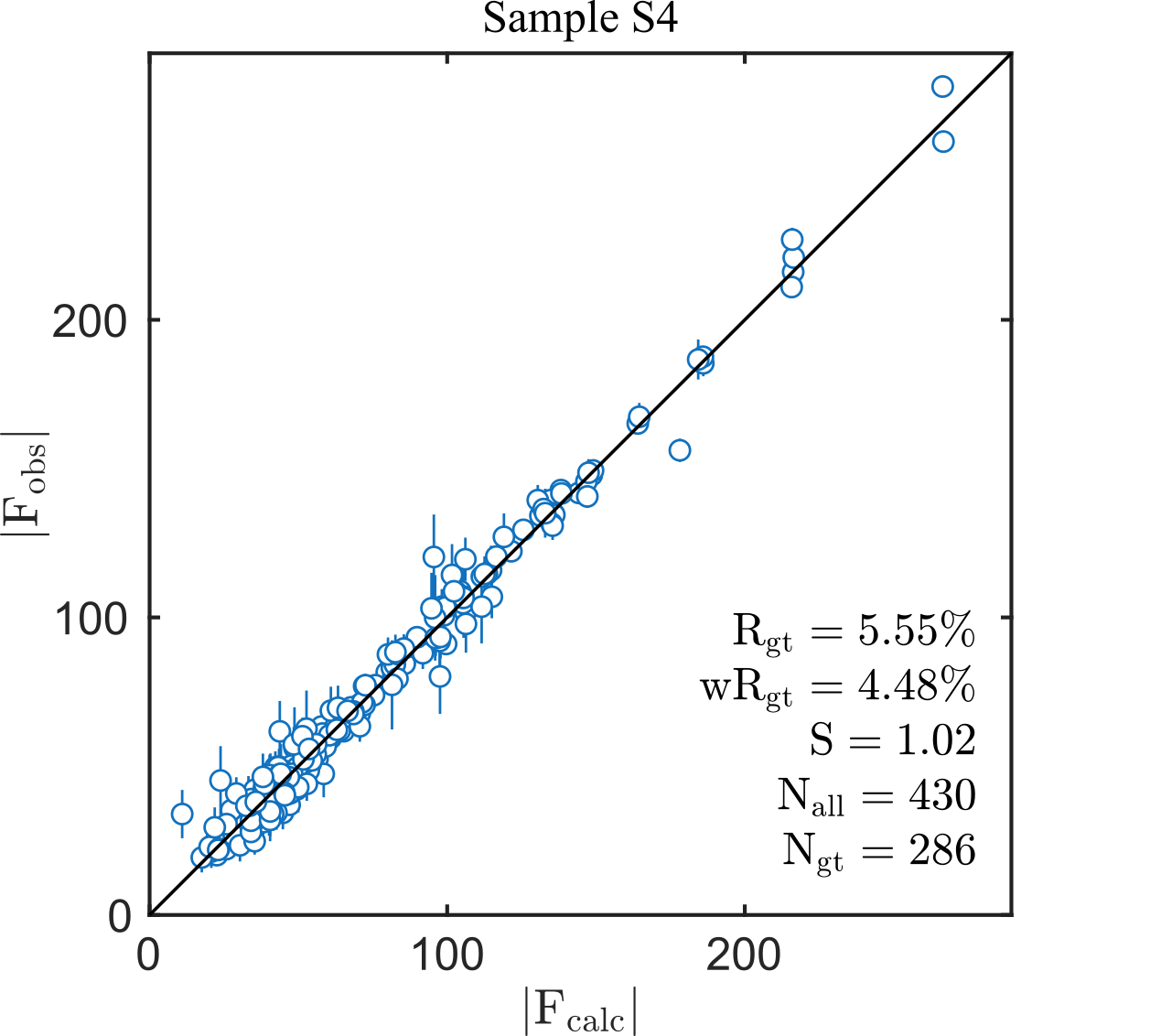}
\caption{
Observed vs. calculated structure factor magnitudes for the diffraction data of un-twinned sample S4 in Fig.~\protect\ref{F:Diff} assuming fixed structural parameters from Table~\protect\ref{T:XRD}. Solid line indicates the 1:1 agreement. The observed structure factors $|F_\text{obs}|$ were divided by an overall fitted scale factor and multiplied by an isotropic extinction factor $\left(1+0.001\xi|F_\text{calc}|^2\lambda^3/\sin2\theta\right)^{1/4}$, where $F_\text{calc}$ is the calculated structure factor, $\lambda=0.71073$~\AA~is the X-ray wavelength, $2\theta$ is the total scattering angle where the reflection is observed and $\xi=0.0113(11)$ is the fitted extinction coefficient. The original list of reflections was reduced in the $2/m$ Laue class to give N$_\text{all}$ symmetry-distinct reflections, out of which N$_\text{gt}$ with $I/\sigma(I)>2$ were included in the fit, with the refinement statistics given in the inset.}
\label{F:1to1}
\end{figure}

Structural characterization of all the samples used for the torque experiments was performed using single-crystal X-ray diffraction with a Mo source SuperNova diffractometer, both at room temperature and at 100~K under cryogenic N$_2$ gas flow. A reference room-temperature crystal structure was obtained by refining the diffraction data of a high-quality, completely un-twinned reference single crystal (sample S4) of $\simeq 50~\mu$m diameter. This confirmed the nominal monoclinic $C2/m$ crystal structure~\cite{Johnson2015} and the refined structural parameters in Table~\ref{T:XRD} are in broad agreement with previous studies~\cite{Stahl2024}. The observed and calculated diffraction patterns are compared in Fig.~\ref{F:Diff} and the quantitative refinement agreement in shown Fig.~\ref{F:1to1}.

\begin{table}
\caption{Structural parameters obtained from refining the room temperature X-ray data of reference sample S4 obtained using the \textsc{Jana2006} package \protect\cite{Jana2006}, with the overall agreement illustrated in Fig.~\protect\ref{F:1to1}: monoclinic space group $C2/m$ with lattice parameters $a=5.9874(8)$~\AA, $b=10.3591(12)$~\AA, $c=6.0424(11)$~\AA, $\beta=108.730(16)^\circ$. Fractional atomic coordinates, isotropic displacement parameters and their uncertainties reflect an average of refinement results for the criterion for reflections included in the refinement $I/\sigma(I)>t$ with the threshold value $t$ varied between 0 and 4. Structural refinement with fully-anisotropic displacement parameters for all atoms also converged, but gave only marginal improvements to the overall agreement (not shown).}
\begin{ruledtabular}
\begin{tabular}{cccccc}
Site&
Wyckoff&
$x$&
$y$&
$z$&
10$^4\!\times U$\!\! (\AA$^2$)\\
\colrule
Ru1& $4g$ & 0 & 0.33340(18) & 0 & 86(12)\\
Cl1& $8j$ & 0.7508(6) & 0.1745(3) & 0.7671(7) & 125(19)\\
Cl2& $4i$ & 0.7279(9) & 0 & 0.2357(11) & 120(20)\\
\end{tabular}
\end{ruledtabular}
\label{T:XRD}
\end{table}

\begin{figure*}[tb]
\centering
\includegraphics[width=\textwidth]{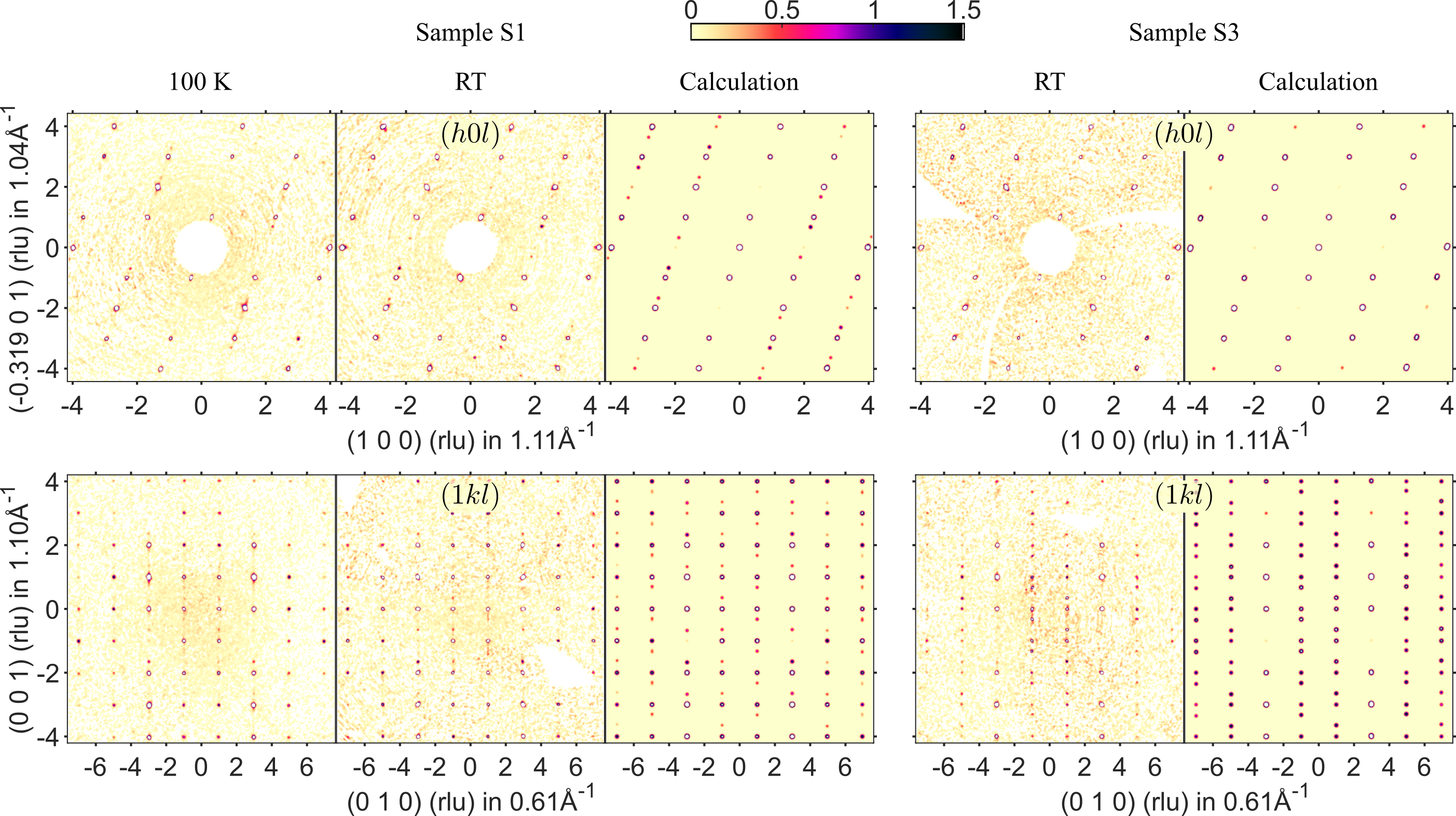}
\caption{Observed X-ray diffraction patterns for sample S1 at 100~K under cryogenic N$_2$ gas flow (column 1) unchanged from room temperature (RT, column 2) and well explained by the nominal monoclinic $C2/m$ structure (column 3) with parameters in Table~\protect\ref{T:XRD}. The small observed diffraction peaks at non-integer $l$ in columns 1 and 2 are attributed to two additional twins rotated by $-60^\circ$ and $120^\circ$ around the $c^*$ axis, of percentage weight 1.6(5)\% and 3.3(5)\%, respectively, included in the calculated patterns, with the quantitative refinement agreement for the room temperature data illustrated in Fig.~\protect\ref{F:1to1_S1}. Columns 4 and 5 show the observed and calculated X-ray diffraction patters for the two-twin sample S3. The twin (53(1)\% of the total mass) is rotated relative to the main grain by $-120^\circ$ around the $c^*$ axis. Top and bottom rows show distinct representative diffraction planes, $(h0l)$ and $(1kl)$, respectively. The color represents diffraction intensity on a log$_{10}$ scale.}
\label{F:All_Diff}
\end{figure*}

\begin{figure}[bt]
\centering
\includegraphics{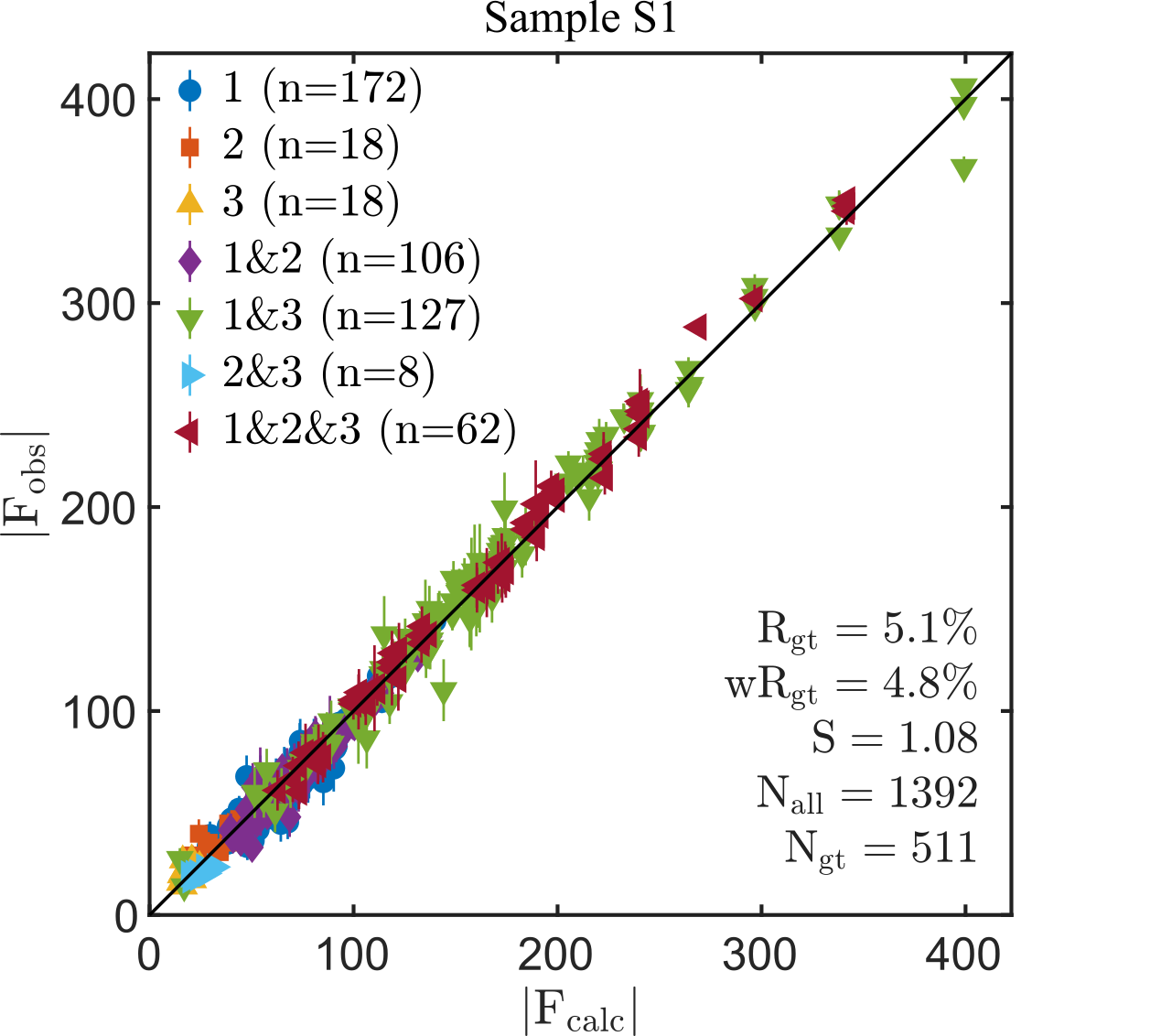}
\caption{Same as Fig.~\protect\ref{F:1to1} but for sample S1 that contained two additional twins, so in total 3 crystal grains co-existed. The legend shows the number of diffraction peaks in distinct groups, single labels (1, 2 or 3) indicate that only one grain contributes, combined labels (1\&2 etc.) indicate that multiple grains contribute to the same diffraction peak. This merged data set for twins was obtained by integrating the raw diffraction data in the reciprocal space of a tripled $a \times b \times 3c $ monoclinic supercell to capture all diffraction peaks from all grains within a common unit cell metric. Each diffraction peak was then assigned to the appropriate group of single or overlapping reflection after converting its Miller indices into the frame of each grain using the twinning rules and assuming an idealized monoclinic cell with pseudo hexagonal metric with $b=\sqrt{3}a$ and $c=-a/(3\cos\beta)$. Intensities were averaged for diffraction peaks belonging to the same group where all contributing reflections were symmetry-equivalent for each grain separately. An overall scale factor and relative twin weights were refined against N$_\text{gt}$ reflections with $I/\sigma(I)>2.4$ assuming fixed structural parameters as per Table~\protect\ref{T:XRD} and extinction coefficient $\xi=0.005$.}
\label{F:1to1_S1}
\end{figure}

Samples S1 and S2 used for the torque experiments were structurally also of high quality with sharp diffraction peaks and no detectable diffuse scattering along $l$ indicating minimal, if any, layer stacking faults, but both samples contained additional twins at the level of a few percent weight. This is illustrated for sample S1 by the diffraction patterns in Fig.~\ref{F:All_Diff} (column 2), where very weak diffraction peaks (the intensity color map is on a log$_{10}$ scale to enhance weak features) are detected near non-integer $l=n\pm1/3$ values ($n$ integer). The complete diffraction pattern including the additional weak peaks is well accounted for (column 3) by including the contribution from the co-existing twins, with the quantitative refinement agreement shown in Fig.~\ref{F:1to1_S1}. The corresponding diffraction data on the same sample at 100~K are shown in Fig.~\ref{F:All_Diff} (column 1) and confirm that the structure is essentially unchanged from room temperature (column 2), apart from small lattice contraction effects. The torque sample S2 had a similar room temperature diffraction pattern to sample S1 and was also over 94(1)\% single grain (not shown). In contrast, Fig.~\ref{F:All_Diff} (column 4) shows diffraction patterns from sample S3, which can be quantitatively explained (column 5) by including the contribution of two almost equal weight twins related by a $120^{\circ}$ rotation around the common $c^*$ axis. Torque measurements for this sample are presented in Appendix~\ref{A:twins}.

\section{Mean-field calculations}\label{A:calculations}
\begin{figure}[tb]
\centering
\includegraphics[scale=1]{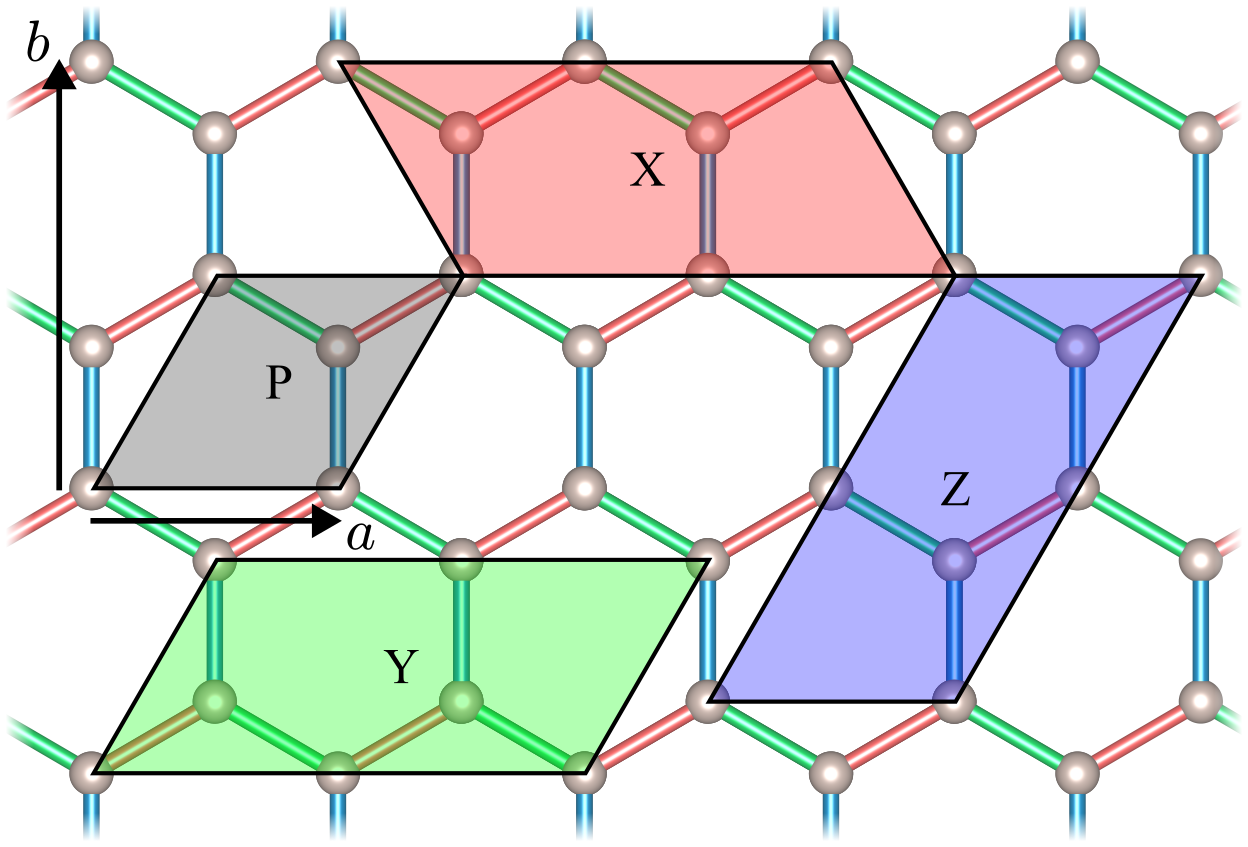}
\caption{Single Ru honeycomb layer, with primitive magnetic cells of the zigzag-X, -Y, -Z and paramagnetic-P phases shaded, color-coded red, green, blue and grey, respectively. Ru bonds are colored as defined in Fig.~\ref{F:C2m}(b).}\label{F:primitives}
\end{figure}

Here we provide technical details of the mean-field calculations to compare with the torque data and the experimental magnetic phase diagrams. The mean-field equilibrium spin configuration of the tested spin Hamiltonians for a given field magnitude and orientation relative to the crystallographic axes 
was obtained using the \textsc{SpinW} package \cite{spinw}. The exchange interactions in the Hamiltonian in Eq.~(\ref{E:H3fold}) are expressed in terms of spin components along cubic Kiteav axes, denoted by SansSerif symbols ${\sf xyz}$ [see Fig.~\ref{F:C2m}(b)], related to the monoclinic crystallographic axes by
\begin{equation}
\bm{a} \parallel \bm{\mathsf{\hat{x}}} + \bm{\mathsf{\hat{y}}}-2\bm{\mathsf{\hat{z}}}, \quad
\bm{b} \parallel \bm{\mathsf{\hat{y}}} - \bm{\mathsf{\hat{x}}}, \quad
\bm{c}^{*} \parallel \bm{\mathsf{\hat{x}}} + \bm{\mathsf{\hat{y}}} + \bm{\mathsf{\hat{z}}},\nonumber
\end{equation}
where the hat $\,\bm{\hat{}}\,$ above the symbols indicates unit vector. In total, four magnetic phases occur depending on the field orientation and magnitude, zigzag-X,-Y,-Z and field-polarized paramagnetic (P). The corresponding magnetic in-plane primitive cells are illustrated by the shaded areas in Fig.~\ref{F:primitives} and their basis vectors are expressed in terms of the monoclinic unit cell basis vectors by
\begin{equation}
\label{E:primitives}
\begin{aligned}
\begin{pmatrix}
\bm a_{\rm X} \\ \bm b_{\rm X}
\end{pmatrix}
&=
\begin{pmatrix}
2 & 0 \\
-\tfrac12 & \tfrac12
\end{pmatrix}
\begin{pmatrix}
\bm a \\ \bm b
\end{pmatrix},
&
\begin{pmatrix}
\bm a_{\rm Y} \\ \bm b_{\rm Y}
\end{pmatrix}
&=
\begin{pmatrix}
2 & 0 \\
\tfrac12 & \tfrac12
\end{pmatrix}
\begin{pmatrix}
\bm a \\ \bm b
\end{pmatrix},
\\[0.6em]
\begin{pmatrix}
\bm a_{\rm Z} \\ \bm b_{\rm Z}
\end{pmatrix}
&=
\begin{pmatrix}
1 & 0 \\
1 & 1
\end{pmatrix}
\begin{pmatrix}
\bm a \\ \bm b
\end{pmatrix},
&
\begin{pmatrix}
\bm a_{\rm P} \\ \bm b_{\rm P}
\end{pmatrix}
&=
\begin{pmatrix}
1 & 0 \\
\tfrac12 & \tfrac12
\end{pmatrix}
\begin{pmatrix}
\bm a \\ \bm b
\end{pmatrix}.
\end{aligned}
\nonumber
\end{equation}
For each applied field, the spin configuration was equilibrated in each of the above four primitive cells, and the one with the lowest energy was selected as the ground state. When one of the X, Y or Z phases was degenerate with the P state, the latter was chosen as the ground state. In each case the equilibration was performed multiple times starting from random spin configurations to avoid being trapped in local energy minima, which could happen in parameter space close to first order phase transition boundaries. This is needed because for the studied Hamiltonian for a general field orientation, field-induced transitions are first order. The torque (per spin) was obtained from the equilibrated spin configuration via $\bm{\tau}=\bm{m}\times\mu_0\bm{H}$ with the magnetic moment (per spin) obtained via a direct summation of the individual $n$ magnetic moments in the primitive magnetic cell, i.e. $\bm{m}=\sum^n_{i=1} \bm{m}_i/n$, where $n=4$ and 2 for any of the three zigzag and P phases, respectively.

\begin{figure*}[tb]
\centering
\includegraphics[scale = 1]{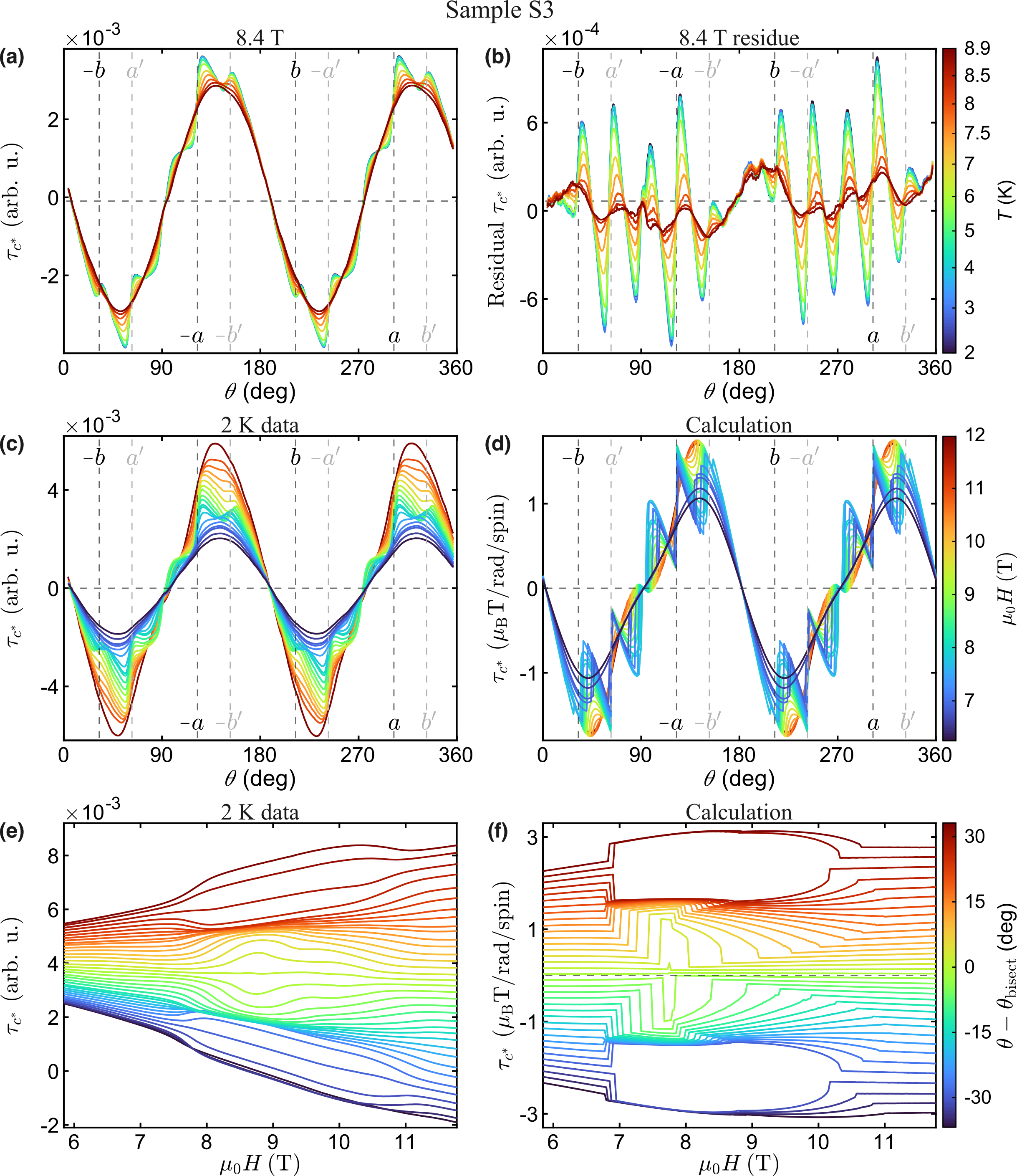}
\caption{Torque for field in the $ab$ plane for the two-twin sample S3. (a) Angular dependence of torque as a function of temperature (color-coded) at 8.4~T. Labeled dashed vertical lines indicate when the applied field is along a reference crystallographic axes, with un-primed/primed black/gray symbols denoting the main grain/twin, respectively. Global stable equilibrium (zero torque with negative gradient) occurs for field along the direction almost bisecting the acute angle between $-b'$ and $b$, which are the stable equilibrium directions for each of the two individual grains of almost equal weight. (b) Residue of the data in (a) after subtracting from each trace a fitted two-fold sinusoid. The revealed multiple anomalies are attributed to the angular scan path crossing phase boundaries in the superimposed magnetic phase diagram of two twins in Fig.~\protect\ref{F:A:phasediagrams_twins}. (c) Torque angular-dependence as a function of field magnitude (color-coded) at 2~K, to be compared with data for an un-twinned sample in Fig.~\protect\ref{F:ab}(a). (e) Torque as a function of field magnitude for a selection of angles (color coded), at 2~K, on both sides of the unstable equilibrium direction $\theta_{\rm bisect}$ that bisects the $-a'$ and $a$ axes (corresponding data for an un-twinned sample is in Fig.~\protect\ref{F:ab}(e)). (d) and (f) show corresponding calculated torque for the same parameters as the data in (c) and (e), respectively, for the minimal monoclinic Hamiltonian assuming equal weight twins with the twin rotated $-120^{\circ}$ around $c^*$ as deduced from the X-ray diffraction data in Fig.~\protect\ref{F:All_Diff} (columns 4 and 5). For clarity, curves in (f) have been vertically offset by $12(\theta-\theta_{\rm bisect})/\theta_{\rm bisect}$.}\label{F:A:ab_twins}
\end{figure*}

\begin{figure}[htb]
\centering
\includegraphics[scale = 1]{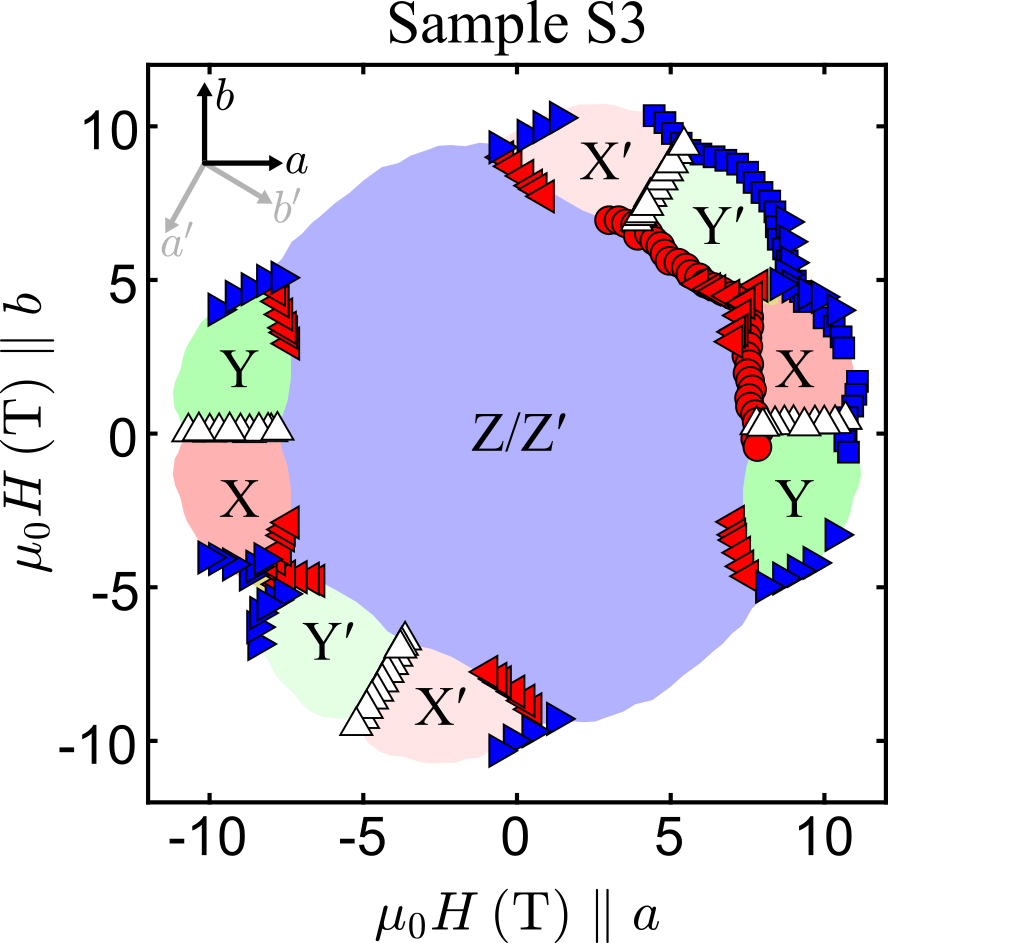}
\caption{Magnetic phase diagram at base temperature (2~K) for field in the $ab$ plane for the two-twin sample S3. Symbols correspond to torque anomalies extracted from the raw torque data in Fig.~\protect\ref{F:A:ab_twins}(c,e), with the symbol shape and color following the same convention as in Fig.~\protect\ref{F:phasediagrams}(a). Shaded regions correspond to distinct magnetic phases in the phase diagram of an un-twinned sample in Fig.~\protect\ref{F:phasediagrams}(a), with lighter colors and primed labels used for the twin rotated by $-120^{\circ}$ around $c^*$. A single shading is used for the overlapping Z/Z$^{\prime}$ regions. The top-left inset shows the relative orientation of the axes of the two twins (faint grey and primed labels for the rotated twin).}
\label{F:A:phasediagrams_twins}
\end{figure}

\section{Torque studies of a two-twin sample}\label{A:twins}
For completeness, we also collected torque data on a sample (S3) that contained two almost equal weight twins rotated by 120$^\circ$ around the common $c^*$ axis, evidenced by the X-ray diffraction shown in Fig.~\ref{F:All_Diff} (column 4) data and 5 calculation. The torque angular dependence for an intermediate field of magnitude 8.4~T applied in the common $ab$ plane of the twins is shown Fig.~\ref{F:A:ab_twins}(a), where curves are color-coded by temperature. Prominent modulations become visible on the torque curves and grow in magnitude upon cooling. The fine structure of those modulations becomes more apparent in Fig.~\ref{F:A:ab_twins}(b), which plots the residue obtained from the raw torque data in (a) after subtracting a fitted two-fold sinusoidal modulation. Those modulations are genuinely of magnetic origin and disappear upon heating. Fig.~\ref{F:A:ab_twins}(c) shows the torque angular dependence at base temperature (2~K) as a function of field magnitude (color-coded). Clear modulations relative to a two-fold sinusoidal shape are observed in the field range 7.2-11~T, most noticeable are sharp jumps in torque when the field direction crosses any of the $\pm a$, $\pm a'$ axes (primed labels indicate the axes of the twin). Additional S-shaped modulations are located at positions offset symmetrically on either side of those jumps. These features resemble the same type of anomalies observed in the torque data from an un-twinned sample in Fig.~\ref{F:ab}(a), but here appear replicated for both constituent twins. Further confirmation of the magnetic origin of these anomalies comes from measurements of the torque as a function of field magnitude. Fig.~\ref{F:A:ab_twins}(e) shows field magnitude scans, curves are color-coded by the field orientation angle $\theta$, spanning a wide range centered at $\theta_{\rm bisect}$ that bisects the $-a'$ and $a$ directions. There are clear anomalies that vary systematically with the field orientation and are symmetric on either side of $\theta_{\rm bisect}$. The sign-reversal in the anomalies near the lower/upper end of the measured angular range correspond to the field direction rotating through the $-a'$ and $a$ axes, respectively. The torque shape near those angles has a very strong resemblance to the measured torque in an un-twinned sample when the field direction is rotated through the $-a$ axis, as shown in Fig.~\ref{F:ab}(e).

The same systematic analysis to locate anomalies in the raw torque curves and their higher order derivatives has been performed (not shown) as in the case of torque data from un-twinned samples S1 and S2 discussed in the main text. The resultant phase diagram that contains the location of all detected anomalies is shown in Fig.~\ref{F:A:phasediagrams_twins}. Symbol shape and color follow the same convention as the phase diagram of un-twinned samples in Fig.~\ref{F:phasediagrams}(a), i.e. color indicates the nature of the associated phase transition irrespective of which twin it corresponds to. The location of all observed anomalies agree very well with the expected phase boundaries assuming the phase diagram is superimposed for the two twins (different color shaded regions correspond to magnetic phases of each of the two twins).

We have also directly compared the measured torque angular dependence in Fig.~\ref{F:A:ab_twins}(c) with corresponding calculations in panels (d) for the minimal monoclinic Hamiltonian model in the main text, including contributions from both twins. All key features in the experimental torque data are well reproduced by the model, including the overall torque sign, the location of stable equilibrium (torque zero with negative gradient) near the bisector of the $-b^{\prime}$ and $b$ axes, and the presence of sharp jumps when the field crosses the $\pm a$ and $\pm a^{\prime}$ axes with S-shape modulations on either side. Corresponding calculations for the measurements of torque as a function of field magnitude in Fig.~\ref{F:A:ab_twins}(e) are shown in panel (f). Again all key features of the data are well reproduced, including the simultaneous change of the sign of the torque anomalies at the lower and higher transition fields when the field direction crosses the $-a^{\prime}$ and $a$ axes. The change in the shape of the torque curves as the field moves away from those positions towards the bisecting angle where torque anomalies again change sign is also captured qualitatively. We conclude that the minimal monoclinic Hamiltonian discussed in the main text can also provide a consistent description of the data from twinned samples, where the observed very complex torque behavior can be rationalized by considering the superposition of the responses from both constituent monoclinic twins.

\bibliography{mRuCl3_reference.bib}
\end{document}